\documentclass[12pt]{article}

\pdfoutput=1

\usepackage{here}
\usepackage{amsmath,amssymb}
\usepackage{bbold}
\usepackage{bm}
\usepackage[dvipdfmx]{graphicx}
\usepackage{graphicx, color}
\usepackage{wrapfig}
\usepackage{cite}
\usepackage{here}
\usepackage{ulem}

\definecolor{amber}{rgb}{1.0, 0.49, 0.0}
\definecolor{tgreen}{rgb}{0.0, 0.5, 0.0}
\definecolor{applegreen}{rgb}{0.55, 0.71, 0.0}
\definecolor{pink}{rgb}{0.94, 0.5, 0.5}
\definecolor{lightblue}{rgb}{0.39, 0.58, 0.93} 
\definecolor{purple}{rgb}{0.59, 0.44, 0.84} 
\definecolor{bluencs}{rgb}{0.0, 0.53, 0.74} 

\usepackage[
colorlinks=true,linkcolor=bluencs,citecolor=tgreen,urlcolor=blue]{hyperref}

\begin{document}
	\title{
		\begin{flushright}
			\ \\*[-50pt] 
			\begin{minipage}{0.22\linewidth}
				\normalsize
				HUPD-2104\\
				\\*[35pt]
			\end{minipage}
		\end{flushright}
		{\Large \bf 
		Electron EDM arising from  modulus $\tau$ \\ 
		 in the supersymmetric modular invariant flavor models
			\\*[15pt]}}

	\author{ 
		\centerline{
			Morimitsu Tanimoto $^{a}\footnote{E-mail address: tanimoto@muse.sc.niigata-u.ac.jp}$ \ \ 
		and \ \ Kei Yamamoto $^{b,c}\footnote{E-mail address: keiy@hiroshima-u.ac.jp}$} \\*[10pt]
		\centerline{
			\begin{minipage}{\linewidth}
				\begin{center}
					$^a${\it \normalsize
					Department of Physics, Niigata University, Niigata 950-2181, Japan}\\*[5pt]
					$^b${\it \normalsize
						Physics Program, Graduate School of Advanced Science and Engineering,
						Hiroshima University, Higashi-Hiroshima 739-8526, Japan} \\*[5pt]
					$^c${\it \normalsize
						Core of Research for the Energetic Universe, Hiroshima University,
						Higashi-Hiroshima 739-8526, Japan} \\*[5pt]
				\end{center}
		\end{minipage}}
		\\*[90pt]}

	\date{
		\centerline{\small \bf Abstract}
		\begin{minipage}{0.9\linewidth}
			\medskip 
			\medskip 
			\small 
			The electric dipole moment (EDM) of electron is studied 
			in the supersymmetric $\rm A_4$ modular invariant theory of flavors
			with CP invariance.
			The CP symmetry of the lepton sector is broken by fixing the modulus $\tau$.
			Lepton mass matrices are completely consistent with 
			observed  lepton masses and  mixing angles in our model.
			In this framework,
			a fixed  $\tau$ also causes the  CP violation in the soft  SUSY breaking terms. The electron EDM arises from
			 the CP non-conserved   soft SUSY breaking terms.
			 The experimental upper bound of the electron EDM
			excludes the SUSY mass scale below  $4$--$6$\,TeV depending
			on five cases of the lepton mass matrices.
			In order to see the effect of CP phase of the modulus $\tau$,
			we  examine the correlation between 
			the electron EDM and the decay rate of the $\mu \rightarrow e \gamma$ decay, which is also predicted by the soft SUSY breaking terms.
			The correlations are clearly predicted 
			 in contrast to  models of the conventional flavor symmetry.
			 The branching ratio is approximately proportional to the square of $|d_e/e|$.
			 	The SUSY mass scale will be   constrained 
			by the future sensitivity  of the electron EDM, 
			$|d_e/e| \simeq 10^{-30}$\,cm. Indeed, 
		it could probe
		the SUSY mass range of  $10$--$20$\,TeV in our model.
			Thus,
			the electron EDM  provides a severe test of the CP violation via the modulus $\tau$ in the supersymmetric modular invariant theory of flavors.			
		\end{minipage}
	}
	
	\begin{titlepage}
		\maketitle
		\thispagestyle{empty}
	\end{titlepage}
	
	\section{Introduction}

The non-Abelian discrete groups have been discussed   to 
challenge  the	flavor problem  of quarks and leptons
in the standard model (SM)
\cite{Altarelli:2010gt,Ishimori:2010au,Ishimori:2012zz,Hernandez:2012ra,King:2013eh,King:2014nza,Tanimoto:2015nfa,King:2017guk,Petcov:2017ggy,Feruglio:2019ktm}.
Indeed,
supersymmetric (SUSY) modular invariant theories give us  an attractive framework to address the flavor symmetry of quarks and leptons
 with non-Abelian discrete groups \cite{Feruglio:2017spp}.
In this approach, the quark and lepton mass matrices are written in terms of modular forms which are   holomorphic functions of  the modulus  $\tau$.
The arbitrary symmetry breaking sector of the conventional models based on  flavor  symmetries is replaced by the moduli space, and then Yukawa couplings are given by modular forms.  

The well-known finite groups $\rm S_3$, $\rm A_4$, $\rm S_4$, and $\rm A_5$
are isomorphic to the finite modular groups 
$\Gamma_N$ for $N=2,3,4,5$, respectively\cite{deAdelhartToorop:2011re}.
The lepton mass matrices have been given successfully  in terms of  $\rm A_4$ modular forms \cite{Feruglio:2017spp}.
Modular invariant flavor models have been also proposed on the $\Gamma_2\simeq \rm S_3$ \cite{Kobayashi:2018vbk},
$\Gamma_4 \simeq \rm S_4$ \cite{Penedo:2018nmg} and  
$\Gamma_5 \simeq \rm A_5$ \cite{Novichkov:2018nkm}.
Based on these modular forms, flavor mixing of quarks and leptons have been discussed intensively in these years.
Phenomenological studies of the lepton flavors have been done
based on  $\rm A_4$ \cite{Criado:2018thu,Kobayashi:2018scp,Ding:2019zxk}, $\rm S_4$ \cite{Novichkov:2018ovf,Kobayashi:2019mna,Wang:2019ovr} and 
$\rm A_5$ \cite{Ding:2019xna}.
A clear prediction of the neutrino mixing angles and the Dirac CP  phase was given in  the  simple lepton mass matrices  with
the  $\rm A_4$ modular symmetry \cite{Kobayashi:2018scp}.
The  Double Covering groups  $\rm T'$~\cite{Liu:2019khw,Chen:2020udk}
and $\rm S_4'$ \cite{Novichkov:2020eep,Liu:2020akv} were also
realized in the modular symmetry.
Furthermore, phenomenological studies have been developed  in many works
\cite{deMedeirosVarzielas:2019cyj,
	Asaka:2019vev,Ding:2020msi,Asaka:2020tmo,Behera:2020sfe,Mishra:2020gxg,deAnda:2018ecu,Kobayashi:2019rzp,Novichkov:2018yse,Kobayashi:2018wkl,Okada:2018yrn,Okada:2019uoy,Nomura:2019jxj, Okada:2019xqk,
	Kariyazono:2019ehj,Nomura:2019yft,Okada:2019lzv,Nomura:2019lnr,Criado:2019tzk,
	King:2019vhv,Gui-JunDing:2019wap,deMedeirosVarzielas:2020kji,Zhang:2019ngf,Nomura:2019xsb,Kobayashi:2019gtp,Lu:2019vgm,Wang:2019xbo,King:2020qaj,Abbas:2020qzc,Okada:2020oxh,Okada:2020dmb,Ding:2020yen,Nomura:2020opk,Nomura:2020cog,Okada:2020rjb,Okada:2020ukr,Nagao:2020azf,Nagao:2020snm,Yao:2020zml,Wang:2020lxk,Abbas:2020vuy,
	Okada:2020brs,Yao:2020qyy,Feruglio:2021dte,King:2021fhl,Chen:2021zty,Novichkov:2021evw,Du:2020ylx,Kobayashi:2021bgy,Ding:2021zbg,Kuranaga:2021ujd}
while theoretical investigations have been also proceeded \cite{Kobayashi:2018bff,Baur:2019kwi,Kobayashi:2019xvz,Nilles:2020kgo,Nilles:2020nnc,Kikuchi:2020nxn,Kobayashi:2020uaj,Kikuchi:2020frp,Ishiguro:2020nuf,Ding:2020zxw,Ishiguro:2020tmo,Kobayashi:2019uyt,Hoshiya:2020hki,Kikuchi:2021ogn,Ding:2021iqp,Nilles:2020tdp,Baur:2020jwc,Nilles:2020gvu,Baur:2020yjl,Baur:2021mtl}.

The supersymmetric modular invariant theory of flavors 
addresses not only the flavor structure of quarks and leptons, but also 
the flavor structure of their superpartners and leads to specific patterns in soft  SUSY breaking terms  
\cite{Du:2020ylx,Kobayashi:2021bgy}.
Soft SUSY breaking terms were studied in several models with 
non-Abelian flavor symmetries \cite{Ko:2007dz,Ishimori:2008ns,Ishimori:2008au,Ishimori:2009ew,
Dimou:2015cmw}.
Such physics can be observed indirectly  in the low energy experiments like  lepton flavor violating (LFV) processes  \cite{Kobayashi:2021bgy}.

The vacuum expectation value (VEV) of the modulus $\tau$ plays an
important role in modular flavor  symmetry, in particular realization of quark and lepton masses and  their mixing angles.
The modulus VEV is fixed as the potential minimum of the modulus potential, so called  the modulus stabilization in modular flavor models  \cite{Kobayashi:2019xvz,Kobayashi:2019uyt,Ishiguro:2020tmo,Kobayashi:2020uaj}.
At such a minimum, the F-term of the modulus $F^\tau$ may be non-vanishing, and leads to SUSY breaking, 
that is the moduli-mediated SUSY breaking \cite{Kaplunovsky:1993rd,Brignole:1993dj,Kobayashi:1994eh,Ibanez:1998rf}.
This specific pattern of soft SUSY breaking terms 
 has been discussed in the LFV \cite{Kobayashi:2021bgy}.
  
  On the other hand,
  the modular invariance has been also studied in the framework of  the generalized CP symmetry \cite{Novichkov:2019sqv},
   which  is the non-trivial CP transformation  in the non-Abelian  discrete flavor symmetry 
   \cite{Ecker:1981wv,Ecker:1983hz,Ecker:1987qp,Neufeld:1987wa,
   	Grimus:1995zi,Grimus:2003yn}.
   A viable CP invariant lepton model  was proposed in the modular $\rm A_4$ 
   symmetry \cite{Okada:2020brs}, in which the CP symmetry is broken
   by fixing $\tau$, that is, the breaking of the modular symmetry
   (see also \cite{Yao:2020qyy}).
  The phenomenological implication of those models
  were studied by focusing 
   the Pontecorvo-Maki-Nakagawa-Sakata (PMNS) mixing angles \cite{Maki:1962mu,Pontecorvo:1967fh} 
   and the CP violating Dirac phase  of  leptons.
In this framework,
  a fixed  $\tau$ also causes the  CP violation in the soft  SUSY breaking terms. 	The electric dipole moments (EDMs) of charged leptons arise
   from the CP non-conserved   soft SUSY breaking terms.
 The current experimental upper bound of the electron EDM, $|d_e/e|\leq 1.1\times 10^{-29}$ cm at 90\% confidence level
 has been reported by the ACME Collaboration \cite{Andreev:2018ayy}, and the future sensitivity is
 expected to reach up to $|d_e/e|\simeq 10^{-30}$\,cm \cite{Kara:2012ay,Griffith}.
  This  future sensitivity
 put forward the theoretical studies of some models
	\cite{Fujiwara:2020unw,Fujiwara:2021vam}.
	
{In our work, we  discuss the electron EDM in the framework of
 the supersymmetric modular invariant theory of flavors.}
We  take the level 3  finite modular groups, 
 $\Gamma_3$ for the flavor symmetry since 
  the property of $\rm A_4$ flavor symmetry has been well known
 \cite{Ma:2001dn,Babu:2002dz,Altarelli:2005yp,Altarelli:2005yx,
 	Shimizu:2011xg,Petcov:2018snn,Kang:2018txu}.
 Indeed,  viable CP invariant lepton models 
  have been investigated linking to the leptogenesis  recently
  \cite{Okada:2021qdf}.
  In this flavor symmetry, we study the electron EDM 
    by  fixing  $\tau$  in the   soft  SUSY breaking term
    while  the observed  lepton masses and PMNS mixing angles
     are completely reproduced.
The SUSY mass scale is also significantly constrained \cite{Kobayashi:2021bgy} by inputting the
observed upper bound of  LFV,
that is  the $\mu \rightarrow e  \gamma$ decay 
\cite{TheMEG:2016wtm}.
In order to see the effect of CP phase in the modulus $\tau$,
 we  examine the correlation between 
the electron EDM and the decay rate of the $\mu \rightarrow e  \gamma$ decay.
The correlation is clearly seen 
 by putting the SUSY mass parameters.
That is contrast to  
the case of the conventional non-abelian discrete flavor symmetric model
\cite{Feruglio:2009hu,Ishimori:2010su,Dimou:2015cmw}. 
Indeed, our mass insertion parameters are obtained  without uncertainty
 once the lepton mass matrices and the SUSY mass scale  are fixed.


The paper is organized as follows.
In section \ref{sec:CP},  we give a brief review on the  CP transformation
in the modular symmetry. 
In section \ref{soft}, we present the soft SUSY breaking terms in the modular flavor models.
In section \ref{A4model},  we present the CP invariant lepton mass matrix in the $\rm A_4$ modular symmetry.
In section \ref{sec:EDM}, we  present formulae for  the electron EDM
and
the branching ratio of the   $\mu \rightarrow e  \gamma$ decay  in terms of the soft SUSY breaking masses.
In section \ref{sec:Num}, we present  the numerical result of the electron EDM as well as the branching ratio of the   $\mu \rightarrow e  \gamma$ decay.
Section \ref{sec:Summary} is devoted to the summary.
In Appendices \ref{app:A4} and \ref{app:modular}, we give  the tensor product  of the $\rm A_4$ group and the modular forms, respectively.
In Appendices \ref{app:RGE} and \ref{app:Loopfun},  we present
the relevant renormalization group equations (RGEs)
and  loop functions, respectively.
In Appendix \ref{app:Me}, we show the charged lepton mass matrix with only 
weight 2 modular forms and corresponding slepton mass matrix.



\section{CP transformation in modular symmetry}
\label{sec:CP}
\subsection{Generalized CP symmetry}

The CP transformation is non-trivial if the non-Abelian  discrete flavor symmetry $G$ is set in the Yukawa sector of a Lagrangian \cite{Grimus:2003yn,Branco:2011zb}.
Let us consider the  chiral superfields.
The CP is a discrete symmetry which involves both Hermitian conjugation of a chiral superfield $\psi(x)$ and inversion of spatial coordinates,
\begin{equation}
\psi(x) \rightarrow {\bf X}_{\bf r}\overline \psi(x_P) \ ,
\label{gCP}
\end{equation}
where $x_P=(t,-{\bf x})$ and ${\bf X_{r}}$ is a unitary transformation
of $\psi(x)$ in the irreducible representation $\bf r$ of the discrete flavor symmetry $G$.  This transformation is called
a generalized CP transformation.
If ${\bf X_{r}}$ is the unit matrix,  the CP transformation is  the trivial one. 
This is the case for the continuous flavor symmetry \cite{Branco:2011zb}.
However, in the framework of the non-Abelian discrete family symmetry,
non-trivial choices of ${\bf X_{r}}$ are  possible.
The unbroken CP transformations of ${\bf X_{r}}$ form the group $H_{CP}$.
Then, ${\bf X_{r}}$ must be consistent with the flavor symmetry transformation,
\begin{equation}
\psi(x) \rightarrow {\rho}_{\bf r}(g)\psi(x) \ , \quad g \in G \ ,
\end{equation}
where ${\rho}_{\bf {r}}(g)$ is the representation matrix for $g$
in the irreducible representation $\bf {r}$. 

The  condition, which has to be respected for consistent implementation of a generalized CP symmetry along with a flavor symmetry,
is given  as follows \cite{Holthausen:2012dk,Feruglio:2012cw,Chen:2014tpa}:
\begin{equation}
{\bf X}_{\bf r} \rho_{\bf r}^*(g) {\bf X}^{-1}_{\bf r}=
{\rho}_{\bf r}(g') 
\ , \qquad g,\, g' \in G \ .
\label{consistency}
\end{equation}
This  is called the consistency condition for ${\bf X}_{\bf r} $.

\subsection{Modular symmetry}
The modular group $\bar\Gamma$ is the group of linear fractional transformations
$\gamma$ acting on the modulus  $\tau$, 
belonging to the upper-half complex plane as:
\begin{equation}\label{eq:tau-SL2Z}
\tau \longrightarrow \gamma\tau= \frac{a\tau + b}{c \tau + d}\ ,~~
{\rm where}~~ a,b,c,d \in \mathbb{Z}~~ {\rm and }~~ ad-bc=1, 
~~ {\rm Im} [\tau]>0 ~ ,
\end{equation}
which is isomorphic to  $PSL(2,\mathbb{Z})=SL(2,\mathbb{Z})/\{\rm I,-I\}$ transformation.
This modular transformation is generated by $S$ and $T$, 
\begin{eqnarray}
S:\tau \longrightarrow -\frac{1}{\tau}\ , \qquad\qquad
T:\tau \longrightarrow \tau + 1\ ,
\label{symmetry}
\end{eqnarray}
which satisfy the following algebraic relations, 
\begin{equation}
S^2 =\mathbb{1}\ , \qquad (ST)^3 =\mathbb{1}\ .
\end{equation}

We introduce the series of groups $\Gamma(N)$, called principal congruence subgroups, where  $N$ is the level $1,2,3,\dots$.
These groups are defined by
\begin{align}
\begin{aligned}
\Gamma(N)= \left \{ 
\begin{pmatrix}
a & b  \\
c & d  
\end{pmatrix} \in SL(2,\mathbb{Z})~ ,
~~
\begin{pmatrix}
a & b  \\
c & d  
\end{pmatrix} =
\begin{pmatrix}
1 & 0  \\
0 & 1  
\end{pmatrix} ~~({\rm mod} N) \right \}
\end{aligned} .
\end{align}
For $N=2$, we define $\bar\Gamma(2)\equiv \Gamma(2)/\{\rm I,-I\}$.
Since the element $\rm -I$ does not belong to $\Gamma(N)$
for $N>2$, we have $\bar\Gamma(N)= \Gamma(N)$.
The quotient groups defined as
$\Gamma_N\equiv \bar \Gamma/\bar \Gamma(N)$
are  finite modular groups.
In these finite groups $\Gamma_N$, $T^N=\mathbb{1}$  is imposed.
The  groups $\Gamma_N$ with $N=2,3,4,5$ are isomorphic to
$\rm S_3$, $\rm A_4$, $\rm S_4$ and $\rm A_5$, respectively \cite{deAdelhartToorop:2011re}.

Modular forms $f_i(\tau)$ of weight $k$ are the holomorphic functions of $\tau$ and transform as
\begin{equation}
f_i(\tau) \longrightarrow (c\tau +d)^k \rho(\gamma)_{ij}f_j( \tau)\, ,
\quad \gamma\in \bar \Gamma\, ,
\label{modularforms}
\end{equation}
under the modular symmetry, where
$\rho(\gamma)_{ij}$ is a unitary matrix under $\Gamma_N$.

Under the modular transformation of Eq.\,(\ref{eq:tau-SL2Z}), chiral superfields $\psi_i$ ($i$ denotes flavors) with weight $-k$
transform as \cite{Ferrara:1989bc},
\begin{equation}
\psi_i\longrightarrow (c\tau +d)^{-k}\rho(\gamma)_{ij}\psi_j\, .
\label{chiralfields}
\end{equation}

We study global SUSY models.
The superpotential which is built from matter fields and modular forms
is assumed to be modular invariant, i.e., to have 
a vanishing modular weight. For given modular forms 
this can be achieved by assigning appropriate
weights to the matter superfields.

The kinetic terms  are  derived from a K\"ahler potential.
The K\"ahler potential of chiral matter fields $\psi_i$ with the modular weight $-k$ is given simply  by 
\begin{equation}
 \frac{1}{[i(\bar\tau - \tau)]^{k}} \sum_i|\psi_i|^2,
\end{equation}
where the superfield and its scalar component are denoted by the same letter, and  $\bar\tau =\tau^*$ after taking VEV of $\tau$.
The canonical form of the kinetic terms  is obtained by 
changing the normalization of parameters \cite{Kobayashi:2018scp}.
The general K\"ahler potential consistent with the modular symmetry possibly contains additional terms \cite{Chen:2019ewa}. However, we consider only the simplest form of
the K\"ahler potential.

For $\Gamma_3\simeq \rm A_4$, the dimension of the linear space 
${\cal M}_k(\Gamma{(3)})$ 
of modular forms of weight $k$ is $k+1$ \cite{Gunning:1962,Schoeneberg:1974,Koblitz:1984}, i.e., there are three linearly 
independent modular forms of the lowest non-trivial weight $2$,
which form a triplet of the $\rm A_4$ group,
${ Y^{(\rm 2)}_{\bf 3}}(\tau)=(Y_1(\tau),\,Y_2(\tau),\, Y_3(\tau))^T$.
These modular forms have been explicitly given \cite{Feruglio:2017spp}  in the  symmetric base of the 
$\rm A_4$ generators  $S$ and $T$ for the triplet representation
(see Appendix \ref{app:A4})  in  Appendix \ref{app:modular}.


\subsection{CP transformation of the modulus $\tau$ and modular multiplets}
The CP transformation in the modular symmetry  was discussed by using the generalized CP symmetry  in Ref. \cite{Novichkov:2019sqv}.
The CP transformation of the modulus $\tau$ is well defined as:
\begin{align}
\tau \xrightarrow{\, CP\, } -\tau^* \, .
\label{tauCPfinal}
\end{align}





The  CP transformation of modular forms were  given in Ref.\cite{Novichkov:2019sqv} as follows.
Define a modular multiplet of the irreducible representation $\bf r$
of $\Gamma_N$   with weight $k$ as $\bf Y^{\rm (k)}_{\bf r}(\tau)$,
which is transformed as:
\begin{align}
\bf Y^{\rm (k)}_{\bf r}(\tau)
\xrightarrow{\, {\rm CP} \, } Y^{\rm (k)}_{\bf r}(-\tau^*) \, ,
\end{align} 
under the  CP transformation.
The complex conjugated CP transformed modular forms
$\bf Y^{\rm (k)*}_{\bf r}(-\tau^*)$ transform almost like the original multiplets
$\bf Y^{\rm (k)}_{\bf r}(\tau)$  under a modular transformation, namely:

\begin{align}
\bf Y^{\rm (k)*}_{\bf r}(-\tau^*) \xrightarrow{\ \gamma \ }
Y^{\rm (k)*}_{\bf r}(-(\gamma\tau)^*) ={\rm (c\tau+d)^k} 
\rho_{\bf {r}}^*({\rm u}(\gamma)) 
Y^{\rm (k)*}_{\bf r}(-\tau^*) \, ,
\end{align}
where $u(\gamma)\equiv CP \gamma CP^{-1}$
\footnote{$u$ acts on the generator as $u(S)=S$ and  $u(T)=T^{-1}$
	\cite{Novichkov:2019sqv}.}
.
Using the consistency condition of Eq.\,(\ref{consistency}),
which gives  ${\bf X_r^T}\rho_{\bf {r}}^*({\rm u}(\gamma)) =\rho_{\bf {r}}(\gamma) {\bf X_r^T}$, we obtain
\begin{align}
\bf X_r^T Y^{\rm (k)*}_{\bf r}(-\tau^*) \xrightarrow{\ \gamma \ }
{\rm (c\tau+d)^k} \rho_{\bf {r}}(\gamma) 
X_r^T Y^{\rm (k)*}_{\bf r}(-\tau^*) \, .
\end{align}
Therefore, if there exists a unique modular multiplet at 
a level $N$, weight $k$ and representation $\bf r$,
which is satisfied for $N=2$--$5$ with weight $2$,
we can express the modular form $\bf Y^{\rm (k)}_{\bf r}(\tau)$ as:
\begin{align}
\bf Y^{\rm (k)}_{\bf r}(\tau)= {\rm \kappa}  X_r^T Y^{\rm (k)*}_{\bf r}(-\tau^*) \, ,
\label{Yproportion}
\end{align}
where $\kappa$ is a  proportional coefficient.
Make $\bf Y^{\rm (k)*}_{\bf r}(-\tau^*)$ by using
Eq.\,(\ref{Yproportion}) and substitute it for
 $\bf Y^{\rm (k)*}_{\bf r}(-\tau^*)$
in  the right hand side of Eq.\,(\ref{Yproportion}).
Then, one obtains  $\bf X_r^* X_r={\rm |\kappa|^2} \mathbb{1}_r $
since  $\bf Y^{\rm (k)}_{\bf r}(-(-\tau^*)^*)=\bf Y^{\rm (k)}_{\bf r}(\tau)$.
Therefore,
the unitary matrix $\bf X_r$ is symmetric one, and $\kappa=e^{i \phi}$
is a phase, which can be absorbed in the normalization of 
modular forms.
Thus, the modular symmetry restricts  $\bf X_r$ being  symmetric.
In conclusion, the CP transformation of modular forms  is given as:
\begin{align}
\bf Y^{\rm (k)}_{\bf r}(\tau)\xrightarrow{\, {\rm CP} \, }
Y^{\rm (k)}_{\bf r}(-\tau^*) =X_r  Y^{\rm (k)*}_{\bf r}(\tau)\, .
\end{align} 
It is also emphasized that $\bf X_r=\mathbb{1}_r$ satisfies the consistency
condition Eq.\,(\ref{consistency})
in a basis that  generators of $S$ and $T$ of $\Gamma_N$ are represented by symmetric matrices
because of 
$ \rho^*_{\bf {r}}(S)=  \rho^\dagger_{\bf {r}}(S)= \rho_{\bf {r}}(S^{-1})=
\rho_{\bf {r}}(S)$ and 
$ \rho^*_{\bf {r}}(T)=  \rho^\dagger_{\bf {r}}(T)= \rho_{\bf {r}}(T^{-1})$.
Our basis  of $A_4$ generators of Eq.\,(\ref{STbase}) is symmetric one
 in Appendix A.

The CP transformations of  chiral superfields and modular multiplets
are summarized as follows:
\begin{align}
\tau \xrightarrow{\, {\rm CP} \, } -\tau^* \, , \qquad
\psi (x)  \xrightarrow{\, {\rm CP} \, } {\bf X_r} \overline \psi (x_P)\, , \qquad
\bf Y^{\rm (k)}_{\bf r}(\tau)\xrightarrow{\, {\rm CP} \, } 
Y^{\rm (k)}_{\bf r}(-\tau^*)  =X_r  Y^{\rm (k)*}_{\bf r}(\tau)\, ,
\label{CPsummary}
\end{align} 
where  $\bf X_r=\mathbb{1}_r$ can be taken  in the basis of symmetric  generators of $S$ and $T$.
We use this CP  transformation of modular forms with
 $\bf X_r=\mathbb{1}_r$ to construct the CP invariant lepton mass matrices in  section \ref{A4model}.


\section{Soft SUSY breaking terms}
\label{soft}

Let us consider the moduli-mediated SUSY breaking \cite{Kaplunovsky:1993rd,Brignole:1993dj,Kobayashi:1994eh,Ibanez:1998rf}.
We present  the soft SUSY breaking terms due to the modulus F-term, using the unit $M_P=1$, where 
$M_P$ denotes the reduced Planck scale.
In supergravity theory,
the action is given by the  K\"ahler potential $K$, 
 the superpotential $W$ and the  gauge kinetic function $f$.
The kinetic terms  are  derived from a K\"ahler potential.

The K\"ahler potential of chiral matter fields $\psi_i$ with the modular weight $-k_i$ is given simply  by 
\begin{equation}
K^{\rm matter} = K_{i \bar i}|\psi_i|^2\,,\quad\qquad K_{i \bar i}= \frac{1}{[i(\bar\tau - \tau)]^{k_i}}\, .
\end{equation}
Then, the  full K\"ahler potential is given as: 
\begin{eqnarray}
K & =& K_0(\tau,M)+
K^{\rm matter} \,, \nonumber \\
K_0(\tau,M) &=& -\ln\,(i(\bar \tau -\tau)) + K(M,\bar M)\,, 
\label{kahler}
\end{eqnarray}
where $M$ denotes moduli other than $\tau$.

The superpotential $W$ is given as:
\begin{eqnarray}
W= Y_{ijk}(\tau)\,\Phi_i \Phi_j \Phi_k  + M_{ij}(\tau)\,\Phi_i \Phi_j\cdots \,.
\label{super}
\end{eqnarray}
We suppose that the gauge kinetic function is independent of the modulus $\tau$,
i.e. $f(M)$ since the modulus $\tau$ does not appear
in the gauge kinetic function  at tree level.

Let us consider the case that the SUSY breaking occurs
by some F-terms of  moduli $X$, $F^X$ $(F^X\not= 0)$.
The canonical form of the kinetic terms  is obtained by 
changing the normalization of parameters.
In the canonical normalization,
the soft masses $\tilde m_i$ and the A-term are given as \cite{Kaplunovsky:1993rd}:
\begin{eqnarray}
\tilde m_i^2= m_{3/2}^2-\sum_X |F^X|^2 \partial_X \partial_{\bar X}\ln K_{i \bar i}\,,
\end{eqnarray}
and 
\begin{eqnarray}
A_{ijk} =A_i+A_j+A_k -\sum_X\frac{F^X}{Y_{ijk}} \partial_X Y_{ijk}\,, \nonumber
\end{eqnarray}
\begin{eqnarray}
A_i = \sum_X F^X \partial_X \ln e^{-K_0/3}K_{i\bar i}\,,
\end{eqnarray}
where $i,\, j$ and $k$ denote flavors.
Here, Yukawa couplings $\tilde Y_{ijk}$ in global SUSY superpotential
are related with Yukawa couplings $ Y_{ijk}$ in the supergravity  superpotential as follows:
\begin{eqnarray}
|\tilde Y_{ijk}|^2=e^{K_0}|Y_{ijk}|^2\,.
\end{eqnarray}
That is, the global SUSY superpotential has vanishing 
modular weight while  the supergravity  superpotential has 
the modular weight $-1$.
Our modular flavor model is studied in global SUSY basis.

Suppose the case of   $X=\tau$. The K\"ahler potential $K$ in Eq.\,(\ref{kahler})
leads to  the soft mass
\begin{eqnarray}
\tilde m_i^2= m_{3/2}^2-k_i \frac{|F^\tau|^2}{(2\,{\rm Im}\,\tau)^2} \,,
\label{smass2}
\end{eqnarray}
where $m_{3/2}$ is the gravitino mass.
It is remarked that $\tilde m_i^2$ becomes  tachyonic
if $k_i|F^\tau|^2/(2{\rm Im}\tau)^2$ is larger than $m_{3/2}^2$. 
Since  $\tilde m_{i}$ should be  at least larger than  ${\cal O}(1)$\,TeV,
Eq.\,(\ref{smass2}) provides a significant constraint
 with our phenomenological discussion.

On the other hand, the A-term is written by 
\begin{eqnarray}
A_{ijk}&=&A_{ijk}^0+A'_{ijk}, \nonumber \\
A_{ijk}^0&=& (1-k_i-k_j-k_k)\frac{F^\tau}{2\,{\rm Im}\,\tau}, \qquad\qquad  A'_{ijk}=\frac{F^\tau}{Y_{ijk}}\frac{dY_{ijk}(\tau)}{d \tau} \, .
\label{Aterm}
\end{eqnarray}
Then, we have the soft mass term  $h_{ijk}=Y_{ijk}A_{ijk}$.
Note that in our convention $\tau$ is dimensionless, and $F^\tau$ has dimension one.
Gaugino masses can be generated by F-terms of other moduli, $F^M$, 
while $F^\tau$ has universal contributions on soft masses and A-terms.

If  we have common weights for three generations in the  modular flavor model,  the soft mass $\tilde m_i$ is flavor blind.
Then,  the left-handed and right-handed
slepton mass matrices $\tilde m_{eLi}$ and $\tilde m_{eRi}$ are universal  as:
\begin{eqnarray}
\tilde m_{eLi}^2=\tilde m^2_{eL0},\qquad\qquad \tilde m^2_{eRi}=\tilde m_{eR0}^2 \,,
\label{massLe}
\end{eqnarray}
that is, they are  proportional to the unit matrix,
which does not contribute the LFV.
This is the case in the previous study of Ref.\cite{Kobayashi:2021bgy}.
However, the  condition of the universal slepton masses is relaxed in  our phenomenological discussion by the assignment of different  weights for the three right-handed charged leptons.
Non-universal slepton mass matrices contribute to  the LFV. 

The first term of $A_{ijk}$ term of Eq.\,(\ref{Aterm}) $A^0_{ijk}$ 
also contributes to the  LFV 
in addition to the second term  $A'_{ijk}$
in the case of different  weights for the three right-handed charged leptons.



\section{CP invariant lepton model in  $\rm A_4$ modular symmetry}
\label{A4model}
\subsection{Lepton mass matrices}
\label{charged lepton}
The CP invariant lepton mass matrices have been proposed
in the $\rm A_4$ modular symmetry  \cite{Okada:2020brs,Okada:2021qdf}.
We adopt those ones in order to discuss the soft SUSY breaking terms.
The three generations of the left-handed lepton doublets are assigned to be an $\rm A_4$ triplet $L$,
and the right-handed charged leptons $e^c$, $\mu ^c$, and $\tau ^c$ are $\rm A_4$ singlets 
$\bf 1$, $\bf 1''$ and $\bf 1'$, respectively. The three generations of the right-handed Majorana neutrinos 
are also  assigned to be an $\rm A_4$ triplet $N^c$
 \cite{Okada:2021qdf}. 
The weight of the superfields of left-handed leptons 
is fixed to be $1$ as a reference value.  The weight of right-handed neutrinos 
is also taken to be 
$1$  in order to give a Dirac neutrino mass matrix
in terms of modular forms of weight $2$. 
On the other hand, weights of the right-handed charged leptons 
$e^c$, $\mu ^c$ and $\tau ^c$ are put $(k_e,\, k_\mu,\, k_\tau)$.
Weights of Higgs fields $H_u$, $H_d$ are  fixed to be $0$.
The representations and weights
for MSSM fields and modular forms of weight $k$  are summarized in Table~\ref{tb:lepton}.

\begin{table}[H]
	\centering
	\begin{tabular}{|c||c|c|c|c|c|c|} \hline
		\rule[14pt]{0pt}{1pt}
		&$L$&$(e^c,\mu^c,\tau^c)$&$N^c$ &$H_u$&$H_d$&
		$ Y_{\bf 3}^{ (k)} $
		\\  \hline\hline 
		\rule[14pt]{0pt}{1pt}
		$SU(2)$&$\bf 2$&$\bf 1$& $\bf 1$ &$\bf 2$&$\bf 2$&$\bf 1$\\
		\rule[14pt]{0pt}{1pt}
		$\rm A_4$&$\bf 3$& \bf (1,\ 1$''$,\ 1$'$)& $\bf 3$ &$\bf 1$&$\bf 1$&$\bf 3$\\
		\rule[14pt]{0pt}{1pt}
		weight & $ 1$ &$(k_e,\ k_\mu,\ k_\tau)$ & $1$ & $0$ & $0$ &  $k$ \\ \hline
	\end{tabular}	
	\caption{ Representations and  weights
		for superfields and  relevant modular forms of weight $k$.
	}
	\label{tb:lepton}
\end{table}
At first, we present  the  neutrino mass matrices.
In Table~\ref{tb:lepton}, the $\rm A_4$ invariant superpotential for the neutrino sector, $w_\nu$, is  given as:
\begin{align}
w_\nu &=w_D+w_N, \nonumber \\
w_D&=\gamma _\nu N^cH_u Y^{ (2)}_{\bf 3}L+
\gamma _\nu 'N^cH_u Y^{ (2)}_{\bf 3}L\, , \nonumber \\
w_N&=\Lambda N^cN^c Y^{ (2)}_{\bf 3}\,,
\label{eq:neutrino}
\end{align}
where $\gamma _\nu $ and  $\gamma _\nu '$ are Yukawa couplings,  and 
$\Lambda$ denotes a right-handed Majorana neutrino mass scale.
By putting $v_u$ for  VEV of the neutral component of $H_u$ 
and taking a triplet $(\nu_e,\, \nu_\mu,\,\nu_\tau)$ for neutrinos,
the Dirac neutrino mass matrix, $M_D$, is obtained as
\begin{align}
M_D=\gamma _\nu v_u\begin{pmatrix}
2Y_1 & (-1+g_D)Y_3 & (-1-g_D)Y_2 \\
(-1-g_D)Y_3 & 2Y_2 & (-1+g_D)Y_1 \\
(-1+g_D)Y_2 & (-1-g_D)Y_1 & 2Y_3\end{pmatrix}_{RL}\,,
\label{MD}
\end{align}
where $g_D=\gamma _\nu '/\gamma _\nu $.
On the other hand the right-handed Majorana neutrino mass matrix, $M_N$ is written as follows:
\begin{align}
M_N=\Lambda\begin{pmatrix}
2Y_1 & -Y_3 & -Y_2 \\
-Y_3 & 2Y_2 & -Y_1 \\
-Y_2 & -Y_1 & 2Y_3\end{pmatrix}_{RR}\,.
\label{MR}
\end{align}
By using the type-I seesaw mechanism, the effective neutrino mass matrix, $M_\nu$ is obtained as
\begin{align}
M_\nu=M_D^{\rm T}M_N^{-1}M_D ~.
\label{seesaw}
\end{align}

We propose the charged lepton mass matrices with minimum number of parameters to reproduce the observed lepton masses and PMNS mixing angles.
Indeed, there are four choices of weights right-handed charged leptons,
those are
($k_e=1,\, k_\mu=1,\, k_\tau=5$), 
($k_e=1,\, k_\mu=3,\, k_\tau=5$), 
($k_e=1,\, k_\mu=1,\, k_\tau=7$) and
($k_e=1,\, k_\mu=3,\, k_\tau=7$) 
labeled as cases A, B, C and D, respectively in our numerical study, as will be discussed later. 
Then,
we need modular forms of weight $2$, $4$, $6$ and $8$,
which are presented in Appendix \ref{app:modular}.

Then, the $\rm A_4$ invariant superpotential of the charged leptons, $w_e$,
by taking into account the modular weights is obtained as 
\begin{align}
w_e&=\alpha_e e^c H_d  Y^{ (2)}_{\bf 3}L+
\beta_e \mu^c H_d  Y^{ (k_\mu+1)}_{\bf 3}L+
\gamma_e \tau^c H_d  Y^{ (k_\tau+1)}_{\bf 3}L+
\gamma_e' \tau^c H_d  Y^{ (k_\tau+1)}_{\bf 3'}L~,
\label{chargedlepton}
\end{align}
where  $\alpha_e$,  $\beta_e$, $\gamma_e$, and $\gamma_e'$
are constant parameters.
Under CP, the superfields transform as:
\begin{align}
e^c \xrightarrow{\,  CP\,}\, X_{\bf 1}^* \,\overline  e^c\, , \quad
\mu^c \xrightarrow{\,  CP\,} X_{\bf 1''}^*\, \overline\mu^c\, , \quad
\tau^c \xrightarrow{\,  CP\,}\, X_{\bf 1'}^* \,\overline \tau^c\, , \quad
L \xrightarrow{\,  CP\,}\, X_{\bf 3} \overline  L\, , \quad
H_d \xrightarrow{\,  CP\,}\,\eta_d\, \overline  H_d\, , 
\end{align} 
and we can take $\eta_d=1$ without loss of generality.
Since the representations of  
$S$ and $T$ are symmetric (see Appendix A), 
we can choose $X_{\bf 3}=\mathbb{1}_{\bf 3}$ 
and $X_{\bf 1}=X_{\bf 1'}=X_{\bf 1''}=\mathbb{1}$
as discussed in Eq.\,(\ref{CPsummary}).

Taking a triplet $(e_L,\, \mu_L,\,\tau_L)$ in the flavor base,
the charged lepton mass matrix $M_E$  is simply written  as:    
\begin{align}
\begin{aligned}
M_e(\tau)=v_d \begin{pmatrix}
\alpha_e & 0 & 0 \\
0 &\beta_e & 0\\
0 & 0 &\gamma_e
\end{pmatrix}
\begin{pmatrix}
Y_1^{(2)}(\tau) & Y_3^{(2)}(\tau) & Y_2^{(2)}(\tau) \\
Y_2^{(m)}(\tau) & Y_1^{(m)}(\tau) & Y_3^{(m)}(\tau) \\
Y_3^{(n)}(\tau)+g_eY_3'^{(n)}(\tau) & Y_2^{(n)}(\tau)+g_eY_2'^{(n)}(\tau) & 
Y_1^{(n)}(\tau)+g_eY_1'^{(n)}(\tau)
\end{pmatrix}  ,
\end{aligned}
\label{ME(2)}
\end{align}
where $m=k_\mu+1$ and  $n=k_\tau+1$ for weights of modular forms
in our case.
The new parameter $g_e$ is defined as $g_e=\gamma _e'/\gamma _e$ and $v_d$ is  VEV of the neutral component of $H_d$.
The coefficients $\alpha_e$, $\beta_e$ and 
$\gamma_e$ are taken to be  real without loss of generality.
Under  CP transformation,  the mass matrix $M_E$ is transformed
following from  Eq.\,(\ref{ME(2)}) as:
\begin{align}
\begin{aligned}
&M_e(\tau)  \xrightarrow{\,  CP\,}  M_e (-\tau^*) = M_e^* (\tau)= \\
&
v_d \begin{pmatrix}
\alpha_e & 0 & 0 \\
0 &\beta_e & 0\\
0 & 0 &\gamma_e
\end{pmatrix}
\begin{pmatrix}
Y_1^{(2)}(\tau)^* & Y_3^{(2)}(\tau)^* & Y_2^{(2)}(\tau)^* \\
Y_2^{(m)}(\tau)^* & Y_1^{(m)}(\tau)^* & Y_3^{(m)}(\tau)^* \\
Y_3^{(n)}(\tau)^*+g_e^*Y_3'^{(n)}(\tau)^* & Y_2^{(n)}(\tau)^*+g_e^*Y_2'^{(n)}(\tau)^* & 
Y_1^{(n)}(\tau)^*+g_e^*Y_1'^{(n)}(\tau)^*
\end{pmatrix}  .
\end{aligned}
\label{CPME}
\end{align}

In a CP conserving modular invariant theory, both CP and modular symmetries are broken spontaneously by  VEV of the modulus $\tau$.
However, there exists certain values of $\tau$ which conserve CP while breaking the modular symmetry.
Obviously, this is the case if
$\tau$ is left invariant by CP, i.e.
\begin{align}
\tau \xrightarrow{\,  CP\,}   -\tau^*=\tau\, \, ,
\label{CPtau}
\end{align}
which indicates $\tau$ lies on the imaginary axis, ${\rm Re} [\tau]=0$.
In addition to ${\rm Re} [\tau]=0$, 
CP is conserved at the boundary of the fundamental domain.

Due to Eq.\,(\ref{CPsummary}),
one then has
\begin{align}
M_\nu(\tau)=M_\nu^*(\tau)\, ,\qquad\qquad M_e(\tau)=M_e^*(\tau) \, ,
\label{CPMassmatrix}
\end{align}
if   $g_{e}$ and  $g_{D}$ are taken to be   real.
Therefore,
the source of the CP violation is only non-trivial ${\rm Re}[\tau]$
after breaking the modular symmetry.
Numerical results of the CP violation 
 have been obtained by
fixing  the modulus $\tau$ with   real $g_{e}$ and  $g_{D}$.

\subsection{Soft masses of sleptons}
As presented in section \ref{soft},
 the SUSY breaking due to the  modulus F term gives 
the soft mass terms of sleptons, $\tilde m^2_{L}$,
$\tilde m^2_{R}$ and $\tilde m^2_{RL}$ as: 
\begin{eqnarray}
&&(\tilde m^2_{eR})_{ii}= m_{3/2}^2-k_i \frac{|F^\tau|^2}{(2\,{\rm Im}\,\tau)^2} \, , \qquad\qquad
(\tilde m^2_{eL})_{jj}= m_{3/2}^2-k_j \frac{|F^\tau|^2}{(2\,{\rm Im}\,\tau)^2} \, , \nonumber\\
&&(\tilde m^2_{eRL})_{ij}\equiv v_d h_{ijk}
=v_d(1-k_i-k_j)\frac{F^\tau}{2\,{\rm Im}\,\tau}Y_{ij} +v_d{F^\tau}\frac{dY_{ij}(\tau)}{d \tau}\, , 
\end{eqnarray}
where $i,\,j$ denote the right-handed and left-handed flavors and 
the subscript index $k$ is omitted in $h_{ijk}$,
and the   weight of Higgs fields $k_k$ in Eq.\,(\ref{Aterm}) is set to be $zero$ without loss of generality.
The subscript indices $L$ and $R$ refer to the chirality of the corresponding SM leptons.
The Yukawa matrix $Y_{ij}$ is given by the charged lepton mass matrix in
Eq.(\ref{ME(2)}) of
 subsection \ref{charged lepton} as  $M_E/v_d$.
 Slepton mass matrices  $\tilde m^2_{eL}$ and 
 $\tilde m^2_{eR}$ are diagonal matrices,
 on the other hand,   $\tilde m^2_{eRL}$ has off-diagonal entries
  in the present flavor basis
  \footnote{
  The SUSY sector of  neutrinos is neglected since the right-handed
 Majorana neutrinos decouples at the high energy scale in our model.
The effect of the right-handed neutrinos is discuss in  section \ref{results}.}.
It is noted that the mass term $\tilde m^2_{eLR}$ is given by
 ${\tilde m_{eRL}^{2\ \dagger}}$.

Let take the models in subsection \ref{charged lepton},
where weights of three right-handed charged leptons
are $k_e$, $k_\mu$ and $k_\tau$, respectively.
On the other hand, $k_j$ of weights for left-handed leptons  are universal as $1$,
because left-handed leptons are constituents of a $\rm A_4$ triplet.

The soft masses of $L$ and $R$ are given:
\begin{align}
&\begin{aligned}
\tilde m^2_{eL}=
\begin{pmatrix}
m_{3/2}^2- |m_F|^2 & 0 & 0 \\
0 &  m_{3/2}^2- |m_F|^2&  0\\
0 & 0& m_{3/2}^2- |m_F|^2 \\
\end{pmatrix}
\end{aligned} \, ,\\
\nonumber\\
&\begin{aligned}
\tilde m^2_{eR}=
\begin{pmatrix}
m_{3/2}^2-k_e |m_F|^2 & 0 & 0 \\
0 &  m_{3/2}^2-k_\mu |m_F|^2&  0\\
0 & 0& m_{3/2}^2-k_\tau |m_F|^2 \\
\end{pmatrix}
\end{aligned} \, ,\\
\end{align}
where
\begin{align}
m_F=\frac{F^\tau}{2\,{\rm Im}\,\tau}\, .
\label{mF}
\end{align}
Thus, $\tilde m^2_{eL}$ matrix is universal 
for flavors (proportional to unit matrix),
but $\tilde m^2_{eR}$ one is not universal in our models.
Therefore,  after moving to  the super-PMNS base
(diagonal base of the neutrino and charged leptons), the off-diagonal entries 
of $\tilde m^2_{eR}$ appear, but   the off-diagonal entries 
of $\tilde m^2_{eL}$ {are not induced}
\footnote{We  neglect RGE effects from Yukawa couplings of leptons since
	 they are very small at $\tan \beta\simeq 5$, which is used in the numerical calculations of section \ref{results}.}
. 

As discussed in  Eq.\,(\ref{smass2}),
 the slepton masses become  tachyonic
if $k_i|F^\tau|^2/(2{\rm Im}\tau)^2$ is larger than $m_{3/2}^2$. 
Therefore,
 the magnitude of $F^\tau$ is significantly constrained
for the larger weight $k_i$ in our phenomenological discussion.

The $\tilde m^2_{eRL}$ matrix has a different flavor structure, which
is shown   as:
\begin{align}
&\tilde m^2_{eRL}\simeq v_d\ \times \nonumber\\
\nonumber\\
&\begin{aligned} 
\left [  m_F
\begin{pmatrix}
-k_e\alpha_e & 0 & 0 \\
0 &-k_\mu\beta_e & 0\\
0 & 0 &-k_\tau\gamma_e
\end{pmatrix} 
\begin{pmatrix}
Y_1^{(2)} & Y_3^{(2)} &  Y_2^{(2)} \\
Y_2^{(m)} &  Y_1^{(m)}&  Y_3^{(m)}\\
Y_3^{(n)}+g_e Y_3^{'(n)} & Y_2^{(n)}+g_e Y_2^{'(n)}& Y_1^{(n)}+g_e Y_1^{'(n)} \\
\end{pmatrix} \right .
\end{aligned}  \nonumber\\
\nonumber\\
&\begin{aligned}
+ \  F^\tau
\left .\begin{pmatrix}
\alpha_e & 0 & 0 \\
0 &\beta_e & 0\\
0 & 0 &\gamma_e
\end{pmatrix} \frac{d}{d \tau}
\begin{pmatrix}
Y_1^{(2)} & Y_3^{(2)} &  Y_2^{(2)} \\
Y_2^{(m)} &  Y_1^{(m)}&  Y_3^{(m)}\\
Y_3^{(n)}+g_e Y_3^{'(n)} & Y_2^{(n)}+g_e Y_2^{'(n)}& Y_1^{(n)}+g_e Y_1^{'(n)} \\
\end{pmatrix} \right ]
\end{aligned}  \, , 
\label{m2RL}
\end{align}
where $m=2$ or $4$, and   $n=6$ or  $8$ for weights of modular forms
 in our models.  The second term of right-hand side in
  Eq.\,({\ref{m2RL}}) is the derivative of the modular forms with respect to the  modulus $\tau$.

  The parameters in these slepton mass matrices, 
$m_{3/2}$ and $F^\tau$ are  taken to be  real
to give the CP conserving modular invariant model.
The CP violation is  caused by fixing $\tau$
  in the soft mass terms as well as in the lepton mass matrices.
   We also suppose real gaugino masses.

In order to study the phenomenological implications of the soft SUSY breaking sector, we  rotate these slepton mass matrices into the physical basis where the Yukawa matrices are real diagonal and positive, i.e. the super-PMNS basis. Any misalignment between the lepton and slepton flavor matrices gives a source of CP violation  and LFV  in the low-energy phenomena.


With these soft masses, the amount of flavor violation can be 
{addressed} in terms of the dimensionless mass insertion parameters.
We adopt the definition 
in Ref.\cite{Dimou:2015cmw} for mass insertion parameters
 because  slepton masses are not universal for flavors.
 The $(i ,j)$ elements of mass insertion parameters are given as: 
\begin{eqnarray}
(\delta_{eLL})_{ij}=\frac{(\tilde m^2_{eL})_{ij}}{\langle \tilde m_e \rangle_{LL}^2}\, ,
\quad 
(\delta_{eRR})_{ij}=\frac{(\tilde m^2_{eR})_{ij}}{\langle \tilde m_e \rangle_{RR}^2}\, ,
\quad (\delta_{eLR})_{ij}=\frac{(\tilde m^2_{ eLR})_{ij}} {\langle \tilde m_{ e} \rangle_{LR}^2}\, ,
\quad 
(\delta_{eRL})_{ij}=\frac{(\tilde m^2_{ eRL})_{ij}}
{\langle \tilde m_{e} \rangle_{RL}^2}\, ,
\end{eqnarray}
where the averaged masses in the denominators are defined by
\begin{eqnarray}
\langle \tilde m_{e} \rangle_{AB}^2
=\sqrt{(\tilde m^2_{eA})_{ii} \, (\tilde m^2_{eB})_{jj}}\, .
\end{eqnarray}
By using these parameters, we discuss the phenomenological implication
 of our   modular invariant models.
\subsection{RGEs effect of sleptons}

Our model of leptons are set at the high energy $Q_0$.
Therefore, we  take into account  the running effects of 
slepton mass matrices at the low energy scale $Q$. 
The renormalization group equations (RGEs)  are 
shown in Appendix \ref{app:RGE}.
Since Yukawa couplings of charged leptons  are small,
the evolutions of off-diagonal elements are dominated by the gauge couplings.
Thus, the largest contributions of the RGEs evolution for off-diagonal elements of A-term  is flavor independent. Then, we can estimate the running effects  by \cite{Martin:1993zk,Martin:1997ns,Ishimori:2009ew}
\begin{eqnarray}
{A}_{e_{ij}} (Q)
\simeq \exp\left[ \frac{-1}{16\pi^2}\int_{Q_0}^{Q} dt
~ \left ( \frac95 g_1^2+3g_2^2 \right )\right ]{A}_{e_{ij}} (m_\text{GUT})
\approx 1.4\times {A}_{e_{ij}} (Q_0),
\end{eqnarray}
where
  $g_{1,2}$ are the gauge couplings of SU(2)$_L\times U(1)_Y$
  and  $t=\ln Q/Q_0$. Numerical coefficient $1.4$ is obtained
by taking $Q_0=10^{16}$\,GeV and $Q=1$\,TeV for a reference.

Therefore,  the  mass term  $\tilde m^2_{eRL}$ is  given  as
\begin{align}
 &\tilde m^2_{eRL}\simeq 1.4\,v_d\ \times \nonumber\\
 \nonumber\\
&\begin{aligned} 
\left [  m_F
\begin{pmatrix}
-k_e\alpha_e & 0 & 0 \\
0 &-k_\mu\beta_e & 0\\
0 & 0 &-k_\tau\gamma_e
\end{pmatrix} 
\begin{pmatrix}
Y_1^{(2)} & Y_3^{(2)} &  Y_2^{(2)} \\
Y_2^{(m)} &  Y_1^{(m)}&  Y_3^{(m)}\\
Y_3^{(n)}+g_e Y_3^{'(n)} & Y_2^{(n)}+g_e Y_2^{'(n)}& Y_1^{(n)}+g_e Y_1^{'(n)} \\
\end{pmatrix} \right .
\end{aligned}  \nonumber\\
\nonumber\\
&\begin{aligned}
+ \  F^\tau
\left .\begin{pmatrix}
\alpha_e & 0 & 0 \\
0 &\beta_e & 0\\
0 & 0 &\gamma_e
\end{pmatrix} \frac{d}{d \tau}
\begin{pmatrix}
Y_1^{(2)} & Y_3^{(2)} &  Y_2^{(2)} \\
Y_2^{(m)} &  Y_1^{(m)}&  Y_3^{(m)}\\
Y_3^{(n)}+g_e Y_3^{'(n)} & Y_2^{(n)}+g_e Y_2^{'(n)}& Y_1^{(n)}+g_e Y_1^{'(n)} \\
\end{pmatrix} \right ]
\end{aligned}  \, , 
\label{Rm2RL}
\end{align}
where $m=2$ or $4$, and   $n=6$ or  $8$ for weights of modular forms.

In the supergravity framework, 
soft masses for all scalar particles have the common scale denoted 
by $m_0$,  and gauginos also have the common scale $M_{1/2}$. 
Therefore, at $Q_0$, we take real masses as:
\begin{eqnarray}
M_1 (Q_0) = M_2 (Q_0) = M_{1/2} \; ,
\end{eqnarray}
where $M_1$ and $M_2$ are the bino and wino masses, respectively.
The effects of RGEs lead at the low energy scale $Q$ 
to  following masses for gauginos \cite{Martin:1993zk,Martin:1997ns}
\begin{eqnarray}
\label{gaugino}
M_1(Q)\simeq\dfrac{\alpha_1(Q)}{\alpha_1(Q_0)}M_1(Q_0),
\qquad\qquad
M_2(Q)\simeq\dfrac{\alpha_2(Q)}{\alpha_2(Q_0)}M_2(Q_0),
\end{eqnarray}
where $\alpha_i=g_i^2/4\pi$ ($i=1,2$) 
and according to the gauge coupling unification at 
$Q_0$, 
$\alpha_1(Q_0)=\alpha_2(Q_0)\simeq 1/25$.
Then, the low energy 
gaugino masses~
\begin{eqnarray}
M_1\approx 0.49\,M_{1/2}\, ,\qquad ~~M_2\approx 0.86\,M_{1/2}\, ,
\label{gauginos}
\end{eqnarray}
by taking $Q_0=10^{16}$\,GeV and $Q=1$\,TeV.

On the other hand,
taking into account the RGEs effect on the average mass scale in
 $\tilde m_{eL}^2$
and $\tilde m_{eR}^2$, we have \cite{Martin:1993zk,Martin:1997ns}
\begin{eqnarray}
\begin{array}{rcl}
\label{smass}
\tilde m_{eL}^2(Q)&\simeq& \tilde m_{eL}^2(Q_0)+K_2(Q)
+\frac{1}{4} K_1(Q)\; ,\\[3mm]
\tilde m_{eR}^2(Q)&\simeq& \tilde m_{eR}^2(Q_0)+K_1(Q) \; ,
\end{array}
\end{eqnarray}
where 
\begin{eqnarray}
K_1(Q)=\frac{3}{5}\,\frac{1}{2\pi^2} \int_{\ln Q}^{\ln Q_0}  dt\, g_1^2(t) M_1(t)^2\, , \qquad 
K_2(Q)=\frac{3}{4}\,\frac{1}{2\pi^2} \int_{\ln Q}^{\ln Q_0}  dt\, g_2^2(t) M_2(t)^2\, .
\end{eqnarray}
We neglect the hyperfine splitting 
${\cal O}(M_Z^2)$ in the slepton mass spectrum produced by electroweak symmetry breaking because of $M_{1/2}\gg M_Z$.
We obtain numerically 
\begin{eqnarray}
\begin{array}{rcl}
(K_1,\,K_2)\simeq (0.14 M^2_{1/2}\,, \ 0.40 M^2_{1/2})
\, ,
\end{array}
\end{eqnarray}
which are flavor independent.
The soft masses of $L$ and $R$ are given as:
\begin{align}
&\begin{aligned}
\tilde m^2_{eL}\simeq 
\begin{pmatrix}
m_0^2- m_F^2 & 0 & 0 \\
0 &  m_0^2- m_F^2&  0\\
0 & 0& m_0^2- m_F^2 \\
\end{pmatrix}
+(0.40 +\frac{0.14}{4})\, M_{1/2}^2
\begin{pmatrix}
1& 0& 0\\
0&1&0\\
0&0&1\\
\end{pmatrix}
\end{aligned} \, ,\nonumber \\
\nonumber\\
&\begin{aligned}
\tilde m^2_{eR}\simeq 
\begin{pmatrix}
m_0^2-k_e m_F^2 & 0 & 0 \\
0 &  m_0^2-k_\mu m_F^2&  0\\
0 & 0& m_0^2-k_\tau m_F^2 \\
\end{pmatrix}
+0.14\, M_{1/2}^2
\begin{pmatrix}
1& 0& 0\\
0&1&0\\
0&0&1\\
\end{pmatrix}
\end{aligned} \, ,
\label{Rm2R}
\end{align}
where $m_{3/2}=m_0$ is put.

The parameter $\mu$ is given through the requirement of the correct 
electroweak symmetry breaking \cite{Martin:1993zk,Martin:1997ns,Feruglio:2009hu,Dimou:2015cmw} :
\begin{eqnarray}
|\mu|^2=\dfrac{\tilde m_{H_d}^2-\tilde m_{H_u}^2\tan^2\beta}{\tan^2\beta-1}-
\dfrac{1}{2}m_Z^2\;.
\label{defmu}
\end{eqnarray}
At the low energy, $|\mu|^2$ turns to 
\cite{Feruglio:2009hu}
\begin{eqnarray}
|\mu|^2\simeq-\dfrac{m_Z^2}{2}+m_0^2\dfrac{1+0.5\tan^2\beta}{\tan^2\beta-1}+M_{1/2}^2\dfrac{0.5+3.5 \tan^2\beta}{\tan^2\beta-1}\;,
\label{defmuSUGRA}
\end{eqnarray}
which is determined by fixing $m_0$, $M_{1/2}$  and $\tan\beta$.
We also take $\mu$ to be real positive.




 Our predictions of the electron EDM and the branching ratio of
 $\mu \to e \gamma$ are given at the $1$\,TeV mass scale.
 The RGE effects of them below $1$\,TeV are induced 
  by  the Yukawa couplings of  charged leptons and gauge couplings
  $g_1$ and $g_2$.
 We can neglect these  RGE effects  
since the Yukawa couplings of  charged leptons are small in our model
and there is no QCD couplings in the one-loop level.  The gauge coupling contributions below $1$\,TeV are   ${\cal O}(1)$\%.
This contribution does not affect our numerical results.

\section{Electron EDM and  $\mu\to e\gamma$ decay}
\label{sec:EDM}
\subsection{Electron EDM}
The current experimental limit for the electric dipole moment of the electron is given by
ACME collaboration \cite{Andreev:2018ayy}:
\begin{equation}
|d_e|~\lesssim~1.1\times10^{-29}\,e\,\text{cm}\, ,
\label{EDMlimit}
\end{equation}
at 90\% confidence level.
 Precise measurements of  the electron EDM are rapidly being updated.
The future sensitivity is
expected to reach up to $|d_e/e|\simeq  10^{-30}$\,cm \cite{Kara:2012ay,Griffith}.
This bound and future sensitivity
can test the framework of  
the supersymmetric modular invariant theory of flavors.
The corresponding EDM formula of leptons is given as \cite{Dimou:2015cmw}:
\begin{eqnarray}
\label{EDM}
\frac{d_e}{e}&=&\frac{\alpha_1}{8\pi}\frac{M_1}{~~\tilde m_e^4}\,
\tilde m_{eL}\, \mathrm{Im}\Big[-(\delta_{eLR})_{11}C_B(\bar x)\,\tilde m_{eR} \nonumber \\
 &+&\Big{\{}(\delta_{eLL})_{1 i}(\delta_{eLR})_{i 1}C'_{B,L}(\bar x)+(\delta_{eLR})_{1i}(\delta_{eRR})_{i1}C'_{B,R}(\bar x)\Big{\}}\,
 \tilde m_{eR_{ii}}  \\
&-&\Big{\{}(\delta_{eLL})_{1 i}(\delta_{eLR})_{ij}(\delta_{eRR})_{j1}+(\delta_{eLR})_{1 j}(\delta_{eRL})_{ji}(\delta_{eLR})_{i1}\Big{\}}
C''_B(\bar x)\,\tilde m_{eR_{jj}}\Big], \nonumber
\end{eqnarray}
where 
$\tilde m_{eL}$ and $\tilde m_{eR}$
are first  mass eigenvalues of  $\tilde m^2_{eL}$ and $\tilde m^2_{eR}$,
respectively,
and $\tilde m_e$ is the averaged mass of $L$ and $R$ as
 $\tilde m_{e}=\sqrt{\tilde m_{eL}\tilde m_{eR}}$. Moreover $\tilde m_{eR_{ii}}$ denotes the i-th
mass eigenvalue of  $\tilde m^2_{eR}$. 
The expression of Eq.\,\eqref{EDM} is  proportional to the bino mass
$M_1$. 
The dimensionless loop functions $C_B(\bar x)$ etc. are presented 
 in Appendix \ref{app:Loopfun}. 
Since our slepton masses of  $L$ and $R$  are not so different
each other, we adopt the approximate formulae 
for $C_B(\bar x)$, $C^{''}_B(\bar x)$ and $C'_{B,L}(\bar x)$ 
 \cite{Dimou:2015cmw} by using  $\bar{x}=(M_1/\tilde m_{e})^2$.


The dominant contribution to the electron EDM comes from 
the first term of the right-hand side, which is the single chirality
flipping diagonal mass insertion $(\delta_{eLR})_{11}$, 
so called flavor-conserving EDM.
Its imaginary part is non-zero due to the 
VEV of the modulus $\tau$, which allows a non-trivial CP phase
 in $\tilde m^2_{eRL}$ to be different from the phases of the charged lepton mass matrix.
The next-to-leading order term is so called flavored EDM \cite{Hisano:2007cz},
 which is related with the FCNC of leptons.
 We will examine both contribution to the electron EDM.

\subsection{Branching ratio of $\mu\to e\gamma$}
Once non-vanishing  off-diagonal elements in the slepton mass matrices are generated, LFV rare decays like $\mu\to e\gamma$ are naturally
induced by the one-loop diagrams with the exchange of gauginos and sleptons
 \cite{Gabbiani:1996hi,Borzumati:1986qx,Altmannshofer:2009ne}.
The branching ratio of $\mu\to e\gamma$ is given as \cite{Dimou:2015cmw}:
\begin{eqnarray}
\nonumber &\,&{\rm BR}(\mu\to e\gamma)~=\alpha_{\rm em}\,\frac{3}{2\pi} \tan^4\theta_W\,M_W^4\,\bar x\,\frac{\mu^2\,\tan^2\beta}{\tilde m_e^6}\times\\
\nonumber &\times&\Bigg(\left|(\delta_{eLL})_{12}\left(-(\delta_{eLR})_{22}
\frac{\tilde m_{eL} \tilde m_{eR}}{\mu\,\tan\beta\,m_\mu}C'_{B,L}+\frac{1}{2}C'_L+C'_2\right)+
(\delta_{eLR})_{12}\frac{\tilde m_{eL}\tilde m_{eR}}{\mu\,\tan\beta\,m_\mu}C_B\right|^2\\
&+&\left|(\delta_{eRR})_{12}\left(-(\delta_{eLR})^*_{22}
\frac{m_{eL}m_{eR}}{\mu\,\tan\beta\,m_\mu}C'_{B,R}-C'_R\right)+
(\delta_{eLR})^*_{21}\frac{\tilde m_{eL}m_{eR}}{\mu\,\tan\beta\,m_\mu}
C_B\right|^2\Bigg)\,,
\label{BRmu}
\end{eqnarray}
where we put  $\sin^2\theta_W=0.231$.
The dimensionless loop functions  are presented 
in Appendix \ref{app:Loopfun}. 
In our model,
 the leading terms come from 
  $(\delta_{eLR})_{12}$ and $(\delta_{eLR})_{21}^*$
   due to the chiral enhancement.
   The next-to-leading ones arise from $(\delta_{eRR})_{{12}}$.
The off-diagonal component $(\delta_{eLL})_{{12}}$ does not come out in our model.


In SUSY models, the branching ratio of $\ell_{i}\rightarrow  \ell_{j} \ell_{k} \bar{\ell}_{k}$ and the conversion rate of $\mu N \to e N$  are	related simply  as \cite{Altmannshofer:2009ne}:
	\begin{align}
	\frac{{\rm BR}(\ell_{i}\rightarrow  \ell_{j} \ell_{k} \bar{\ell}_{k})}{{\rm BR}(\ell_{i}\rightarrow \ell_{j}\gamma)} &=
	\frac{\alpha_{\rm em}}{3 \pi} \left( 2 \log\frac{m_{\ell_i}}{m_{\ell_k}} - 3\right)\,,\qquad\qquad
	\frac{{\rm CR}(\mu N\to e N)}{{\rm BR}(\ell_{i}\rightarrow \ell_{j}\gamma)} = \alpha_{\rm em}\, ,
	\label{CR}
	\end{align}
	where $\alpha_{\rm em}$ is the electromagnetic fine-structure constant. 
	
Current experimental bounds and future prospects of EDM, $\mu \to e \gamma$ and relevant processes are
summarized  in Table \ref{tb:bound}.
\begin{table}[H]
	\centering
	\scalebox{1}[1]{
		\begin{tabular}{c|c|c} 
			\rule[15pt]{0pt}{0pt}
				&Current bounds &Future prospects\\  \hline\hline 
			\rule[15pt]{0pt}{0pt}
			$|d_e/e|$\,cm 
			&$1.1\times 10^{-29}$  \cite{Andreev:2018ayy}
			&$\sim 10^{-30}$ \cite{Kara:2012ay,Griffith}   \\ 
			\rule[15pt]{0pt}{0pt}
			${\rm BR}(\mu\to e\gamma)$
			&$4.2 \times 10^{-13}$ \cite{TheMEG:2016wtm, Zyla:2020zbs}
			&$6\times 10^{-14}$ \cite{Baldini:2018nnn}   \\
			\rule[15pt]{0pt}{0pt}
			${\rm BR}(\mu\rightarrow   e e \bar e)$
			&$1.0\times 10^{-12}$   \cite{Zyla:2020zbs}
			&$\sim 10^{-16}$  \cite{Blondel:2013ia,Abrams:2012er, Adamov:2018vin} \\ 
			\rule[15pt]{0pt}{0pt}
			${\rm CR}(\mu N\to e N)$
			&$7.0\times 10^{-13}$  \cite{Zyla:2020zbs}
			&$\sim 10^{-16}$    \cite{Blondel:2013ia,Abrams:2012er, Adamov:2018vin}\\ 
			\hline
		\end{tabular}
	}
	\caption{
	Current experimental bounds and future prospects of relevant processes.
	}
	\label{tb:bound}
\end{table}

\section{Numerical results}
\label{sec:Num}
\label{results}
As discussed in subsection \ref{charged lepton}, the CP invariant lepton mass matrices have been given
in the $\rm A_4$ modular symmetry \cite{Okada:2021qdf}.
The tiny neutrino masses are obtained by type-I seesaw.
The CP symmetry is broken spontaneously by the  VEV of the modulus $\tau$.
Thus,  the fixed value of $\tau$ breaks the CP symmetry as well as the modular invariance.
The source of the CP phase is  the real part of $\tau$.
Lepton mass matrices of four cases  of  weights ($k_e,\, k_\mu,\, k_\tau$)
are completely consistent with observed  lepton masses and  PMNS mixing angles.
Then, the CP violating Dirac phase 
is predicted clearly at the fixed value of  $\tau$ \cite{Okada:2021qdf}.
The predicted  CP phases  of five cases are different
as seen in Table \ref{sample}.

By using those successful charged lepton mass matrices,
 we calculate  the electron EDM and the branching ratio
of $\mu \to e \gamma$.
In our numerical analyses, 
we  take four cases A, B, C and D with 
 weights of right-handed charged leptons $(k_e,\, k_\mu,\, k_\tau)$: 
 \begin{eqnarray}
 (k_e,\, k_\mu,\, k_\tau)\,:\  A\,(1,\,1,\, 5) \,,
 \quad B\,(1,\,3,\, 5)\,,
  \quad C\,(1,\,1,\, 7)\,,  \quad D\,(1,\,3,\, 7)\,.
 \nonumber
 \end{eqnarray}
We also  discuss the alternative case E, where the charged lepton mass matrix is written
 with only weight 2 modular forms,
 $ (k_e,\, k_\mu,\, k_\tau)=(1,\,1,\, 1)$ (see Appendix \ref{app:Me}),
  in which the branching ratio
  of $\mu \to e \gamma$ was studied in Ref. \cite{Kobayashi:2021bgy}.
  In the case of E, the neutrino mass matrix is given by
   the dimension-five Weinberg operator instead of type-I seesaw
   \footnote{In the case of  the charged lepton mass matrix 
   	with only weight 2 modular forms, a simple neutrino seesaw mass matrix is not obtained for the model of spontaneously CP violation.}
   .

\begin{table}[H]
	\centering
	\begin{tabular}{|c|c|c|c|c|} \hline 
		\rule[14pt]{0pt}{0pt}
		($k_e,\, k_\mu,\, k_\tau$)&A\,($1,\, 1,\, 5$)&B\,($1,\, 3,\, 5$) &C\,($1,\, 1,\, 7$)&D\,($1,\, 3,\, 7$)\\  \hline
		\rule[14pt]{0pt}{0pt}
		$\tau$&  $  -0.1912 + 1.1194   \, i$ & $0.1931 + 1.1240  \, i$ &$0.0901 + 1.0047 \, i$&
		-0.1027 + 1.0050\, i \\ 
		\rule[14pt]{0pt}{0pt}
		$g_D$ &$ -0.800$ & $ -0.800$ & $-0.660$& $0.685$\\
		\rule[14pt]{0pt}{0pt}
		$g_e$  &  $-0.905$&  $-0.900$& $-0.530$& $-0.573$ \\
		\rule[14pt]{0pt}{0pt}
		$\beta_e/\alpha_e$ & $3.70\times 10^{-3}$& $4.73\times 10^{-3}$ &$5.94\times 10^{-3}$&
		$6.30\times 10^{-3}$ \\
		\rule[14pt]{0pt}{0pt} 
	$\gamma_e/\alpha_e$ &  $ 9.71$ &$ 10.1$ &$17.6$&$16.0$\\ \hline
		\rule[14pt]{0pt}{0pt}
		$\sin^2\theta_{12}$ & $ 0.305$&$ 0.309$ & $0.324$& $0.326$\\
		\rule[14pt]{0pt}{0pt}
		$\sin^2\theta_{23}$ &  $ 0.569$&$ 0.574$ &$0.441$& $0.479$\\
		\rule[14pt]{0pt}{0pt}
		$\sin^2\theta_{13}$ &  $ 0.0222$& $ 0.0225$&$0.0222$&$0.0223$	\\ \hline
		\rule[14pt]{0pt}{0pt}
		$\delta_{CP}^\ell$ &  $172^\circ$&$187^\circ$ & $183^\circ$& $176^\circ$\\
		\rule[14pt]{0pt}{0pt}
		$\sum m_i$ &  $62.5$\,meV &$62.8$\,meV	& $60.5$\,meV & $60.7$\,meV\\
		\rule[14pt]{0pt}{0pt}
		$\sqrt{\chi^2}$ & $1.08$ & $1.43$ & $2.16$ & $2.38$\\
		\hline
	\end{tabular}
	\caption{Numerical values of parameters
		$\tau$, $g_D$, $g_e$, $\beta_e/\alpha_e$,
		$\gamma_e/\alpha_e$  and output of best fitting
		 three mixing angles.
		The CP violating  Dirac phase $\delta_{CP}^\ell$ and
		 the sum of three neutrino masses $\sum m_i$ are predicted.
		 The square root of the sum of $\chi^2$ are also shown.
	}
	\label{sample}
\end{table}
In Table \ref{sample}, we show five  parameters of models,
out put of three mixing angles and 
predicted  the  CP violating Dirac phase and the sum of neutrino masses at the {best-fit values} 
for cases A, B, C and  D
in the normal hierarchy of neutrino masses (NH)
\footnote{There are no allowed parameter set within $3 \,\sigma$
	confidence level for the inverted hierarchy of neutrino masses.}
, where  neutrino data of  NuFit 5.0 are put \cite{Esteban:2020cvm}.
The charged lepton masses are fitted at the high energy scale
$2\times 10^{16}$\,GeV
with $\tan\beta\equiv v_u/v_d=5$ \cite{Antusch:2013jca,Bjorkeroth:2015ora}.
	The parameter $g_D$ appears in the neutrino Dirac mass matrix in Eq.\.(\ref{MD}) and is real as seen in Eq.\,(\ref{CPMassmatrix}).
	Since we discuss  sleptons, which are the  superpartner of the charged lepton sector,
	$g_D$ does not affect our numerical results of the electron EDM
	and the LFV.
	On the other hand, real parameters $g_e$, $\beta_e/\alpha_e$,
	$\gamma_e/\alpha_e$ in addition to complex value of $\tau$ determine the  charged lepton mass matrix.
	By using those four real parameters $g_D$, $g_e$, $\beta_e/\alpha_e$,
	$\gamma_e/\alpha_e$
	and one complex parameters (${\rm Re}\,[\tau]$ and ${\rm Im}\,[\tau]$),
	we performed $\chi^2$-fit, where
	we adopted the sum of one-dimensional 
	$\chi^2$ function for four accurately known dimensionless observables,
	the ratio of two neutrino mass squared differences 
	$\Delta m_{\rm atm}^2/\Delta m_{\rm sol}^2$,
	$\sin^2\theta_{12}$, $\sin^2\theta_{23}$ and $\sin^2\theta_{13}$
	in NuFit 5.0 \cite{Esteban:2020cvm}. In addition, we employed Gaussian approximations for fitting the charged lepton mass ratios,
	$m_e/m_\tau$ and  $m_\mu/m_\tau$.
	Since free six parameters fit six observables,  we can predict 
	the CP violating Dirac phase $\delta_{CP}^\ell$
	and the sum of neutrino masses $\sum m_i$.

Parameters of the case E are shown in Appendix \ref{app:Me}.


We fix the charged lepton mass matrices
by using 
 $\beta_e/\alpha_e$, $\gamma_e/\alpha_e$ and  $g_e$
 in addition to  the modulus $\tau$
in Table \ref{sample} apart from the normalization of the mass matrix.
Then, the mass insertion parameters are determined  including  CP phases
if the  SUSY  parameters $m_{3/2}=m_0$, $M_{1/2}$ and $F^\tau$
are put.
In order to calculate the electron EDM and the branching ratio
of $\mu \to e \gamma$, the slepton mass matrices
of Eqs.\,(\ref{Rm2RL}) and (\ref{Rm2R})
are rotated  into the physical basis where the charged lepton mass matrix is real diagonal and positive.

	As discussed in  Eqs.\,(\ref{smass2}) and (\ref{mF}),
	the magnitude of $F^\tau$ should be  significantly constrained
	for the larger weight $k_i$ to prevent the tachyonic slepton.
	Since the largest weight is $k_\tau=7$,
	 we take $|F^\tau|=m_0/2$ in the following numerical analyses.
	 We will discuss later if $|F^\tau|$ is set to be larger than $=m_0/2$
	 with keeping the slepton mass of ${\cal O}(1)$\,TeV.

\begin{figure}[H]
	\begin{minipage}[]{0.49\linewidth}
		\centering
			\includegraphics[width=0.9\linewidth]{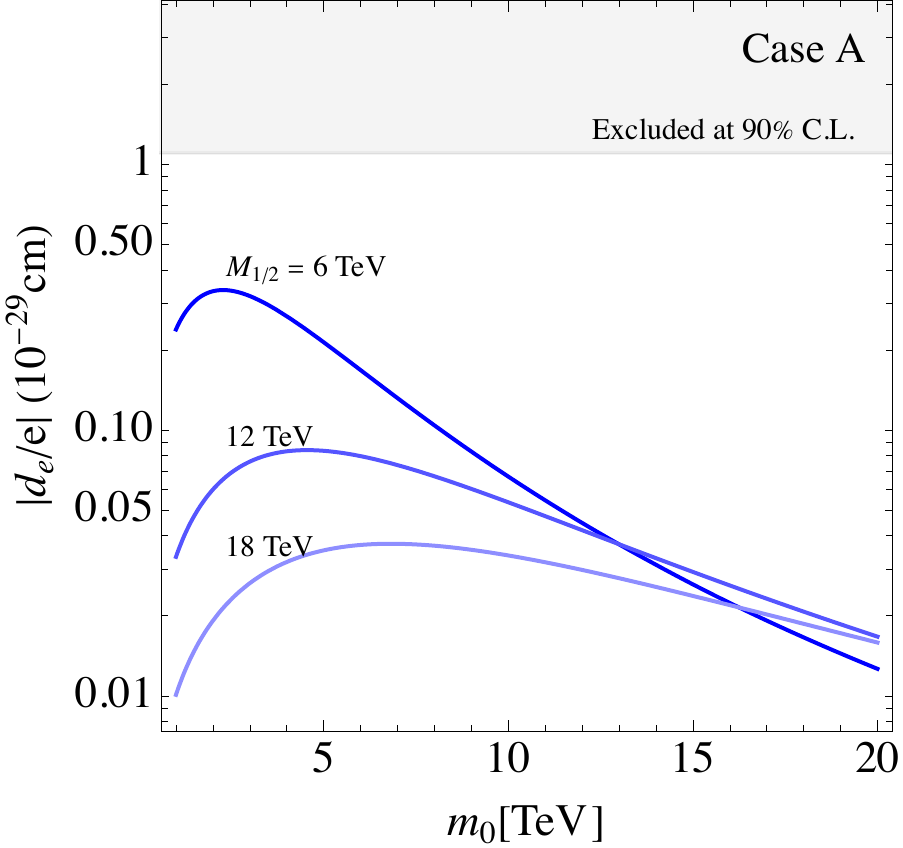} \vspace{-0mm}
		\caption{The predicted $|d_e/e|$ 
			versus $m_0$ with putting $F^\tau=m_0/2$ in case A.	
		The grey region is excluded by the experimental upper bound.}
	\label{fig-edm-1}
	\end{minipage}
	\hspace{2.5mm}
	\begin{minipage}[]{0.49\linewidth}
		\includegraphics[width=0.9\linewidth]{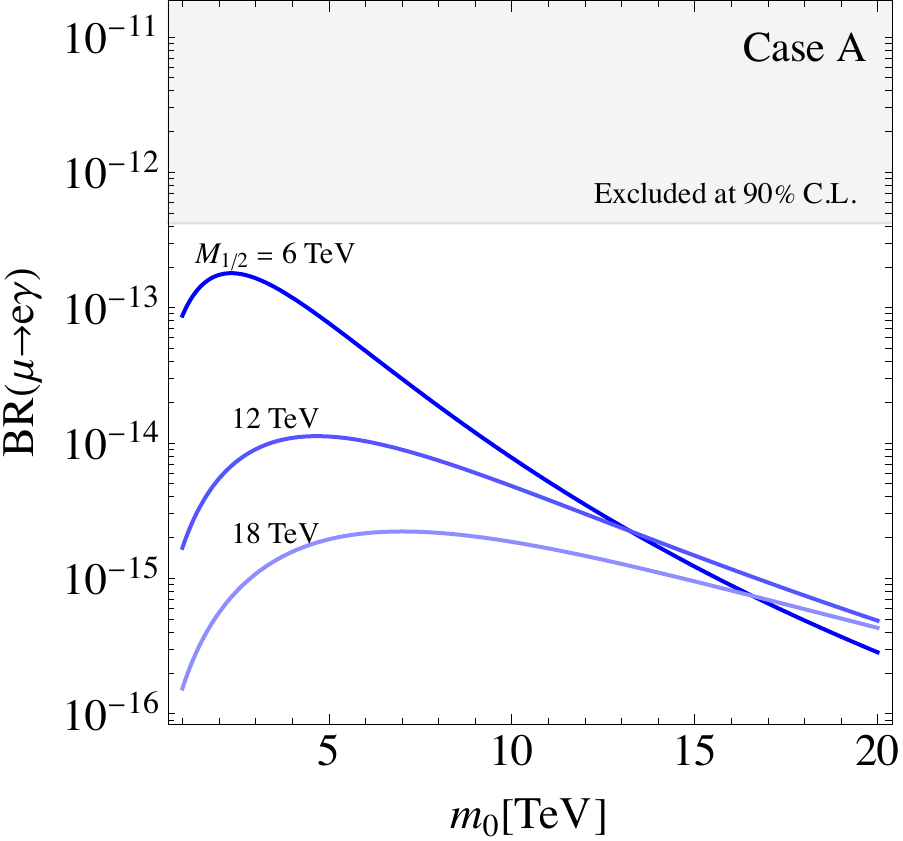} \vspace{-0mm}
		\caption{ The predicted ${\rm BR}(\mu\to e \gamma)$ 
			versus $m_0$ with putting $F^\tau=m_0/2$ in case  A.	
			The grey region is excluded by the experimental upper bound.}
		\label{fig-mue-1}
	\end{minipage}
\end{figure}

At first,
we present our numerical results of the electron EDM, $|d_e/e|$
for  case A.
The SUSY mass parameters are variable in  $M_{1/2}=6$--$18$\,TeV and  $m_0=1$--$20$\,TeV,
which are allowed in the slepton, bino and wino  searches of the LHC experiments \cite{Sarkar:2021lju,Kalogeropoulos:2020ovp}.
We plot $|d_e/e|$ versus $m_0$ in Fig.\,\ref{fig-edm-1}, 
where $F^\tau=m_0/2$ is put.
Three curved lines correspond to $M_{1/2}=6,\,12,\,18$\, TeV,
respectively.
As seen in  Fig.\,\ref{fig-edm-1}, the predicted electron EDM is lower than
the experimental upper bound as far as  the SUSY mass scale is larger than a few TeV for  $F^\tau=m_0/2$.
Indeed, 
the predicted value is consistent with 
the experimental upper bound if  the gaugino mass scale  $M_{1/2}$ is larger than $4$\,TeV.

It would be helpful to comment on the behavior of the predicted curves.
	The maximum values are apparently found at  the low $m_0$ region.
	The predicted values increase at $m_0$ close to $1$\,TeV.
	This behavior is due to taking $F^\tau=m_0/2$ in order to reduce the number of free parameters,
	although $\tilde m_{eRL}^2$ ($A$-term)  is proportional to $F^\tau$
	 and independent $m_0$
	as seen in Eq.\,(\ref{Rm2RL}).
	If $F^\tau$ is fixed to, for example, $1$\,TeV,
	the prediction becomes a monotonically decreasing function against $m_0$.

The SUSY mass scale is also significantly constrained by 
the experimental upper bound of the branching ratio for
the $\mu \rightarrow e  \gamma$ decay  \cite{Kobayashi:2021bgy} .
We  plot  ${\rm BR}(\mu\to e\gamma)$ versus $m_0$ in Fig.\,\ref{fig-mue-1}, 
where we put $F^\tau=m_0/2$ again.
It is found that  the predicted ${\rm BR}(\mu\to e\gamma)$  is lower than
the experimental upper bound as far as 
the gaugino mass scale  $M_{1/2}$ is larger than $6$\,TeV.
Thus, the  $\mu\to e\gamma$ process constrains more severely
the SUSY mass scale  compared with the electron EDM for case A.

\begin{figure}[h]
	\begin{minipage}[]{0.49\linewidth}
		\centering
		\includegraphics[width=0.9\linewidth]{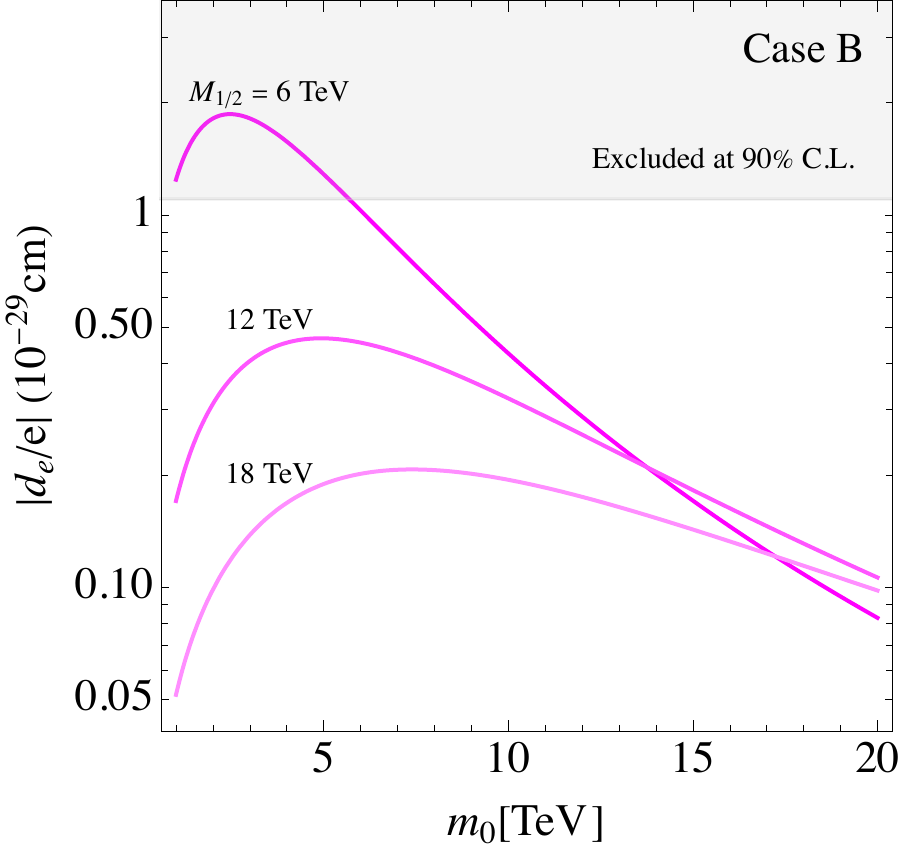} \vspace{-0mm}
		\caption{The predicted  $|d_e/e|$ 
			versus $m_0$ with putting $F^\tau=m_0/2$ in case B.	
			The grey region is excluded by the experimental upper bound.}
		\label{fig-edm-2}
	\end{minipage}
	\hspace{2.5mm}
	\begin{minipage}[]{0.49\linewidth}
		\vspace{0mm}
		\includegraphics[width=0.9\linewidth]{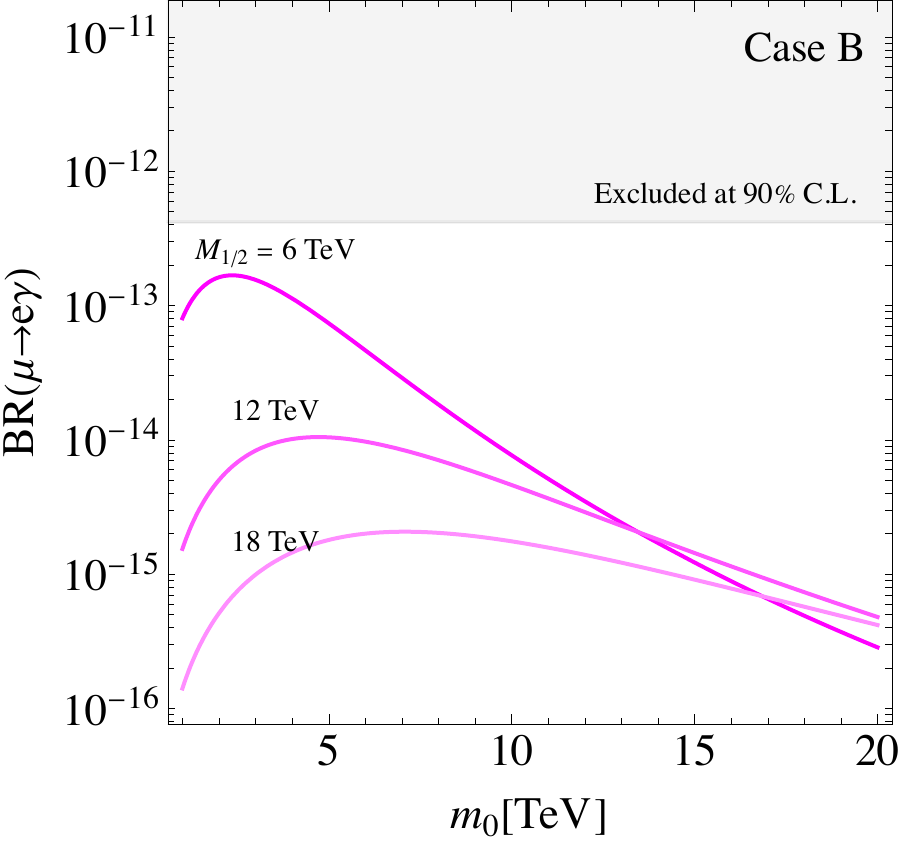}
		\caption{ The predicted  ${\rm BR}(\mu\to e \gamma)$ 
			versus $m_0$ with putting  $F^\tau=m_0/2$ in case B.	
			The grey region is excluded by the experimental upper bound.}
		\label{fig-mue-2}
	\end{minipage}
\end{figure}
In contrast to case A,
the electron EDM constrains tightly
the SUSY mass scale  in case B.
We  present our numerical results of the electron EDM, $|d_e/e|$
and  ${\rm BR}(\mu\to e\gamma)$ for case B.
We plot $|d_e/e|$ versus $m_0$ in Fig.\,\ref{fig-edm-2},
where $F^\tau=m_0/2$ is put.
It is found that  the predicted electron EDM  exceeds
the experimental upper bound
 at $m_0\leq 5$\,TeV  if  the gaugino mass scale  $M_{1/2}$ is  $6$\,TeV.
 
 We  show  ${\rm BR}(\mu\to e\gamma)$ versus $m_0$
with putting $F^\tau=m_0/2$ in Fig.\,\ref{fig-mue-2}.
The predicted ${\rm BR}(\mu\to e\gamma)$  is almost same
 as the one in case  A of Fig. \ref{fig-mue-1}.

Thus, the constraints of the SUSY mass scale 
 from the upper bounds  $|d_e/e|$ and  ${\rm BR}(\mu\to e\gamma)$
 depend on the model of the charged leptons.


\begin{figure}[h]
	\begin{minipage}[]{0.49\linewidth}
		\centering
		\includegraphics[width=0.9\linewidth]{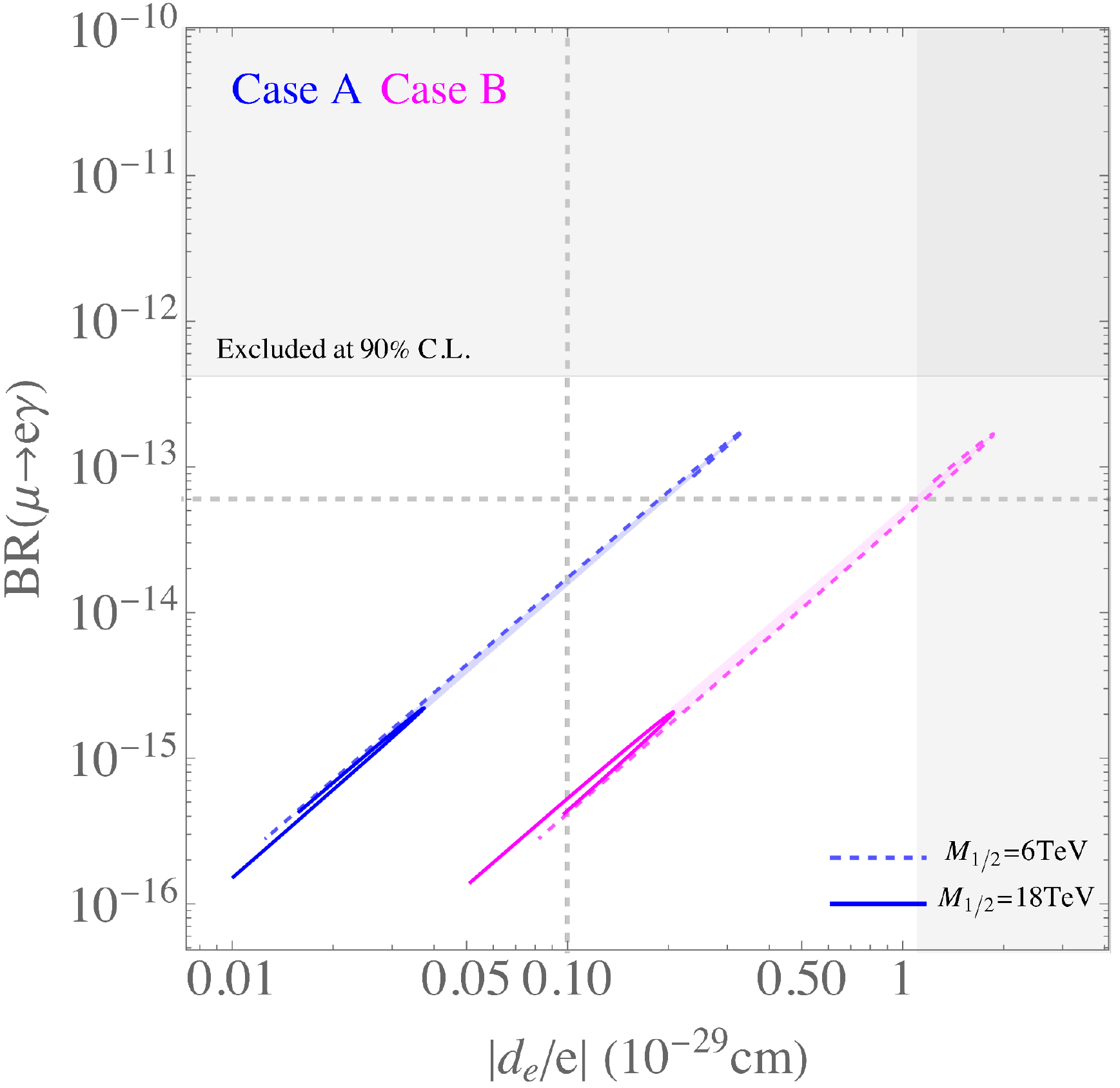} \vspace{-0mm}		
		\caption{Plot of $|d_e/e|$
			and   ${\rm BR}(\mu\to e \gamma)$ for cases A and B with $F^\tau=m_0/2$, where $m_0=1$--$20$\,TeV for fixed $M_{1/2}=6$ and $18$\,TeV.
			The blue curves denotes the predictions at fixed  gaugino mass   $M_{1/2}$ for  case A,  and the pink one for case B.
			The grey regions are excluded by the experimental upper bounds, 
			and the vertical  and horizontal dashed grey  lines indicate the future sensitivity.
		}
		\label{edm-mue-ABm0}
	\end{minipage}
	\hspace{2.5mm}
	\begin{minipage}[]{0.49\linewidth}
		\includegraphics[width=0.9\linewidth]{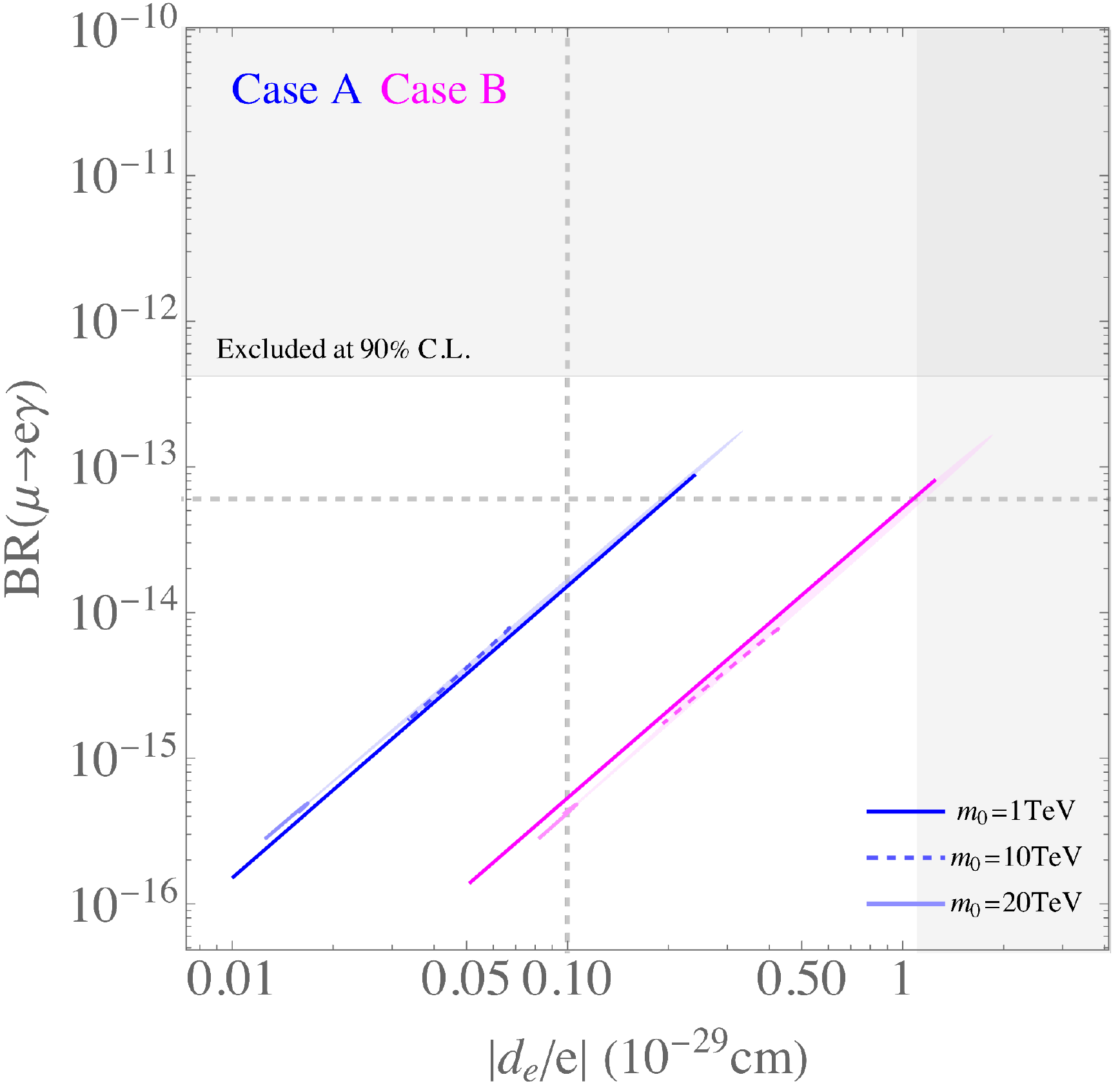}
		\caption{ Plot of $|d_e/e|$
			and   ${\rm BR}(\mu\to e \gamma)$ for cases A and B with $F^\tau=m_0/2$, where $M_{1/2}=6$--$18$\,TeV for  fixed $m_0=1\,,10$ and $20$\,TeV.
				The blue curves denotes the predictions at fixed     $m_0$ for  case A,  and the pink one for case B.
			The grey regions are excluded by the experimental upper bounds, 
			and the vertical  and horizontal dashed grey  lines indicate the future sensitivity.}
		\label{edm-mue-ABM12}
	\end{minipage}
\end{figure}

\begin{figure}[h]
	\begin{minipage}[]{0.49\linewidth}
		\centering
		\includegraphics[width=0.9\linewidth]{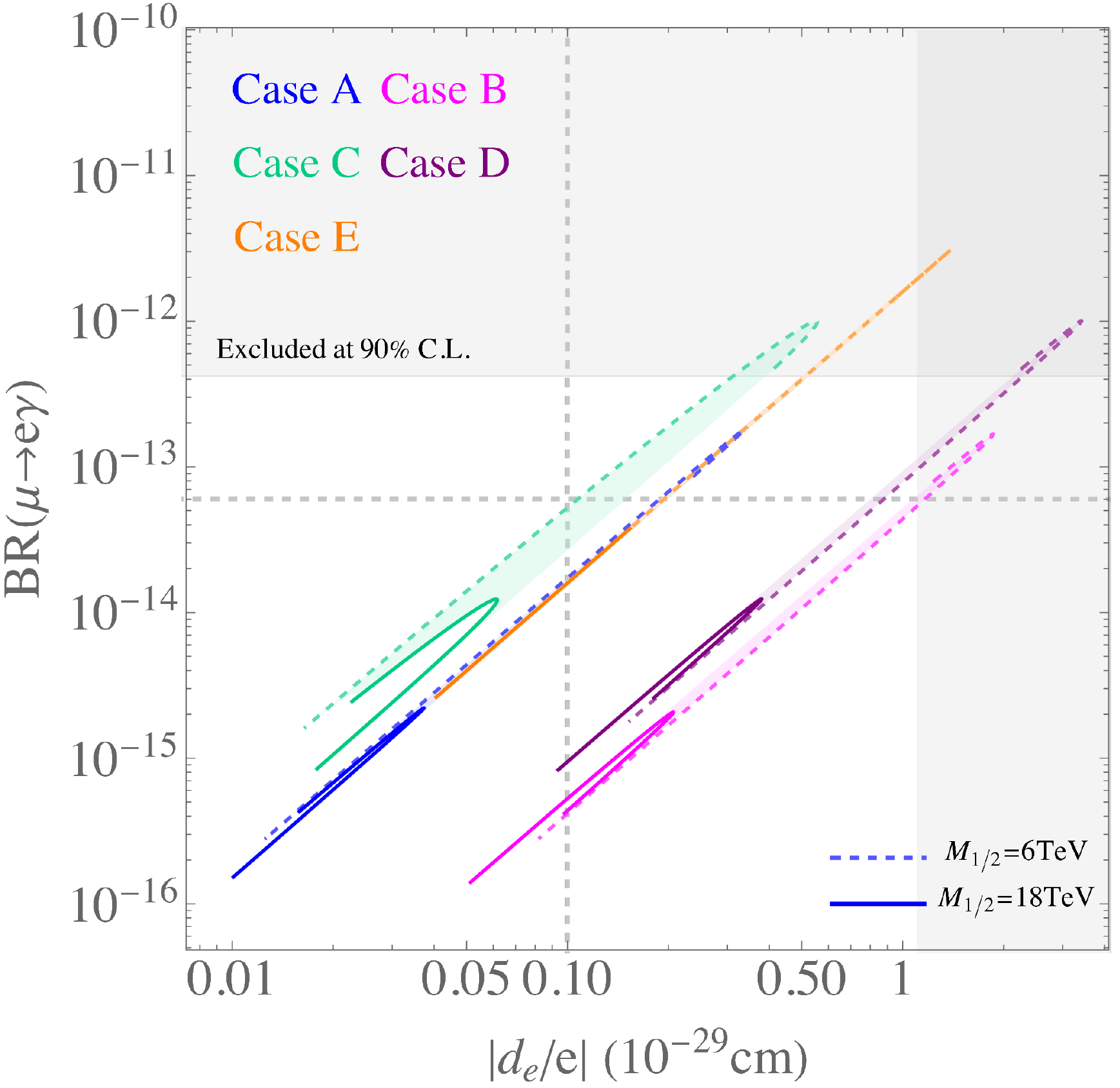} \vspace{-0mm}		
		\caption{Plot of $|d_e/e|$
			and   ${\rm BR}(\mu\to e \gamma)$ for all cases with $F^\tau=m_0/2$, where $m_0=1$--$20$\,TeV for  fixed $M_{1/2}=6$ and $18$\,TeV.
				The blue curves denotes the predictions at fixed  gaugino mass   $M_{1/2}$ for  case A, and the same for other cases.
			The grey regions are excluded by the experimental upper bounds, 
			and the vertical  and horizontal dashed grey  lines indicate the future sensitivity.
		}
		\label{edm-mue-allm0}
	\end{minipage}
	\hspace{2.5mm}
	\begin{minipage}[]{0.49\linewidth}
		\includegraphics[width=0.9\linewidth]{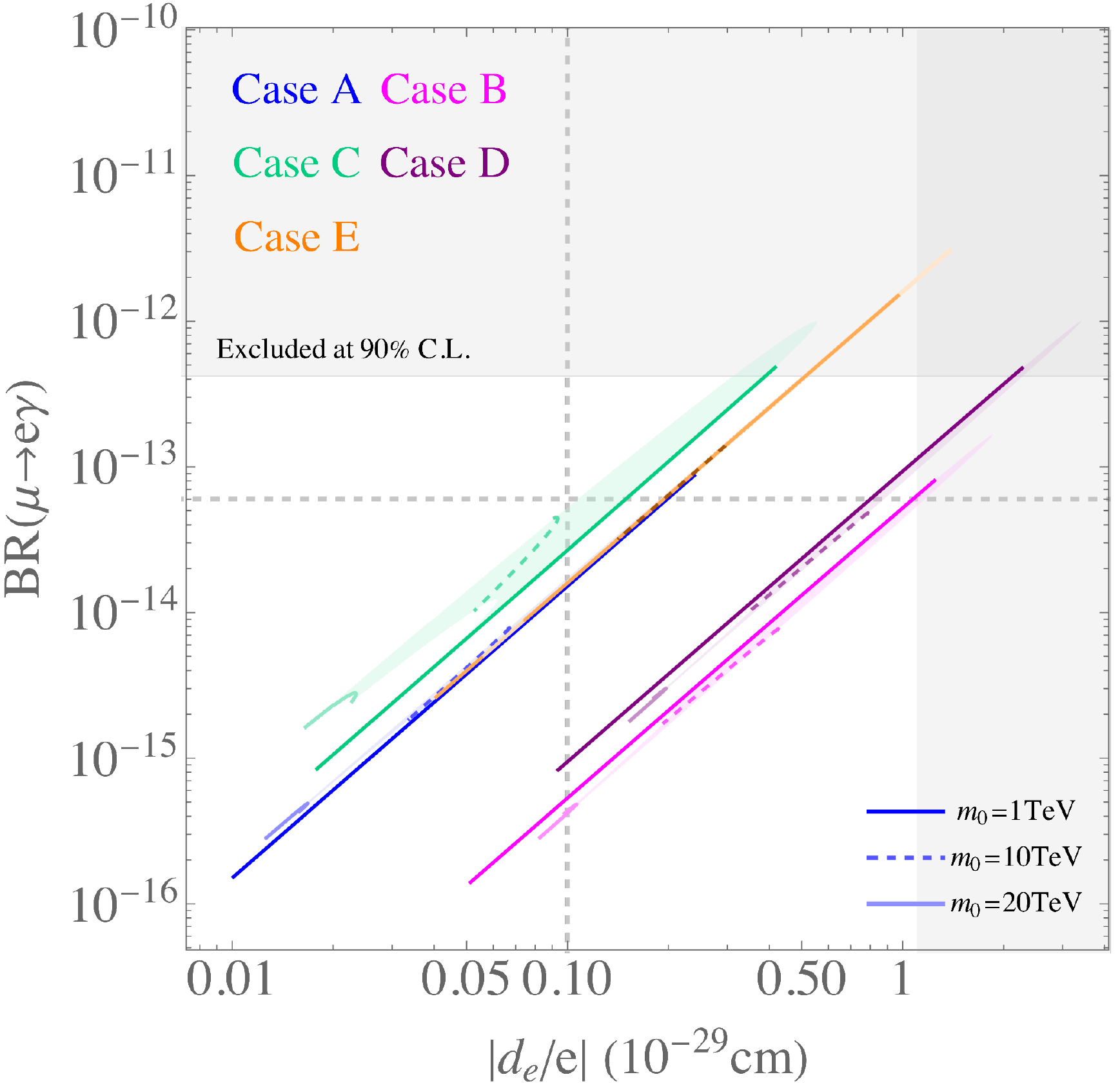}
		\caption{Plot of $|d_e/e|$
				and   ${\rm BR}(\mu\to e \gamma)$ for all cases with $F^\tau=m_0/2$, where $M_{1/2}=6$--$18$\,TeV for fixed $m_0=1\,,10$ and $20$\,TeV.
				The blue curves denotes the predictions at fixed     $m_0$ for  case A,  and the same for other cases.
				The grey regions are excluded by the experimental upper bounds, 
				and the vertical  and horizontal dashed grey  lines indicate the future sensitivity.}
		\label{edm-mue-allM12}
	\end{minipage}
\end{figure}
In order to see the importance of CP phases via the modulus $\tau$,
we  examine the correlation between 
the electron EDM and the decay rate of the $\mu \rightarrow e  \gamma$ decay for both cases A and B. 
The correlation is clearly seen  in Figs.\,\ref{edm-mue-ABm0}
 and \ref{edm-mue-ABM12}.
 We plot them in the range of 
 $m_0=1$--$20$\,TeV with fixing  $M_{1/2}=6$ and $18$\,TeV
  in  Fig.\,\ref{edm-mue-ABm0}, 
 on the other hand, in the range of 
 $M_{1/2}=6$--$18$\,TeV with fixing  $m_0=1$, $10$ and $20$\,TeV
 in  Fig.\,\ref{edm-mue-ABM12}. 

We find the linear correspondence between ${\rm BR}(\mu\to e \gamma)$
and   $|d_e/e|$ in the  logarithmic coordinates
 for both cases A and B.
 The branching ratio is approximately proportional to the square of $|d_e/e|$.
 The slope of the line is independent of the value of  $F^\tau$, although $F^\tau=m_0/2$ is taken in these figures.
 Similar  correlations are  also  found in other cases B--E.
 This provides a crucial test for our predictions in future.
It is also seen that
the   predicted decay rate of  $\mu \rightarrow e  \gamma$ 
is almost same  for both  cases A and B while 
the predicted $|d_e/e|$ of case B  is larger than the one of case A
in factor $5$. 
Thus,  the magnitude of the predicted electron EDM depends on the 
charged lepton mass matrix considerably.

 Although the  bound of the electron EDM 
is $|d_e/e|\leq 1.1\times 10^{-29}$\,cm \cite{Andreev:2018ayy},  the future sensitivity is
expected to reach up to $|d_e/e|\simeq 10^{-30}$\,cm \cite{Kara:2012ay,Griffith}, which is denoted by the vertical dashed grey  line.
For case A,  their sensitive mass scale  is   much below $10$\,TeV
for $m_0$ and $M_{1/2}$ as seen in Figs.\,\ref{edm-mue-ABm0}
and \ref{edm-mue-ABM12}.
On the other hand,
for case B, the electron EDM can probe  the  mass scale $m_0$
of $10$--$20$\,TeV as seen  in Figs.\,\ref{edm-mue-ABm0}
and \ref{edm-mue-ABM12}.
The future sensitivity of the branching ratio ${\rm BR}(\mu\to e \gamma)$ is $6\times 10^{-14}$ \cite{Baldini:2018nnn}, 
which is shown by the horizontal dashed grey line,
 excludes only the SUSY mass region much below $10$\,TeV.

Let us discuss  the model dependence among A--E.
 We show  the correlations between $|d_e/e|$
and   ${\rm BR}(\mu\to e \gamma)$ for
all cases A--E in  Figs.\,\ref{edm-mue-allm0} and \ref{edm-mue-allM12}.
 We plot them in the range of 
$m_0=1$--$20$\,TeV with fixing  $M_{1/2}=6$ and $18$\,TeV
in  Fig.\,\ref{edm-mue-allm0}, 
on the other hand, in the range of 
$M_{1/2}=6$--$18$\,TeV with fixing  $m_0=1$, $10$ and $20$\,TeV
in  Fig.\,\ref{edm-mue-allM12}. 
The predicted $|d_e/e|$ and ${\rm BR}(\mu\to e \gamma)$
of  case A and case  C are similar each other,  
 and  those of case B and case  D are also similar each other
 as seen in  Figs.\,\ref{edm-mue-allm0} and \ref{edm-mue-allM12}.
 The prediction of case E overlaps somewhat  with the one of case A, but its region of case E extends upward considerably.
 The $\mu\to e\gamma$ process of case E constrains most severely
 the SUSY mass scale among A--E. 
 On the other hand,  the electron EDM of case D constrains most
 severely the SUSY mass scale.
 
 The future sensitivity of the electron EDM, $|d_e/e|\simeq 10^{-30}$\,cm \cite{Kara:2012ay,Griffith}
  will probe  the SUSY mass scale of $10$--$20$\,TeV for B, D and E.
On the other hand, the future sensitivity of  the branching ratio ${\rm BR}(\mu\to e \gamma)$,  $6\times 10^{-14}$ \cite{Baldini:2018nnn}
can probe the SUSY mass scale close to  $10$\,TeV only for case E.
For cases A--D,  their sensitive mass scale  is   much below $10$\,TeV.

 It is noted that
  the branching ratio of $\mu\to 3e$ and the conversion rate of $\mu N \to e N$  will be  sensitive for proving the SUSY mass scale of 
  higher than $10$\,TeV
  although the predicted branching ratio and conversion rate 
 are significantly below the current experimental upper bounds
 as discussed later \cite{Blondel:2013ia,Abrams:2012er, Adamov:2018vin}.

\begin{table}[t]
	\centering
	\scalebox{0.89}[0.9]{
		\begin{tabular}{|c|c|c|c|c|c|} \hline
			\rule[15pt]{0pt}{0pt}
			Cases	&A&B&C&D	&E\\  \hline\hline 
			\rule[15pt]{0pt}{0pt}
			$(\delta_{eLR})_{11} \times 10^8$ 
			&$0.46-0.56\, i$ 
			&$1.8+ 3.0\, i$ 
			&$-12-1.0\, i$
			&$-28-5.6 \,i$ 
			&$-56-2.2\, i$ \\
			\rule[15pt]{0pt}{0pt}
			$(\delta_{eLR})_{22} \times 10^5$ 
			&$-1.74+0.42\, i$ 
			&$1.0+0.42 \, i$ 
			&$-1.3-0.39\, i$
			&$0.62-0.37 \,i$ 
			&$-0.15-0.95\, i$ \\
			\rule[15pt]{0pt}{0pt}
			$(\delta_{eLR})_{12} \times 10^{6}$			
			&$0.91-4.42\, i$ 
			&$-0.83-4.2\, i$ 
			&$11-0.60\, i$  
			&$-11-1.0 \,i$ 
			&$7.4+18\, i$ \\
			\rule[15pt]{0pt}{0pt}
			$(\delta_{eLR})_{13}\times10^{4}$			
			&$0.46+0.54\, i$ 
			&$-0.45+0.55 \,i$ 
			&$1.5+4.3\, i$
			&$1.4-3.8\, i$ 
			&$0.15-1.9\, i$ \\
			\rule[15pt]{0pt}{0pt}
			$(\delta_{eLR})_{21}\times10^{7}$			
			&$-0.21+0.04\, i$ 
			&$-1.7-0.12\, i$ 
			&$0.70-0.13 \,i$ 
			&$1.0+0.2 \,i$ 
			&$0.28-0.87\, i$ \\
			\rule[15pt]{0pt}{0pt}
			$(\delta_{eLR})_{23}\times10^{5}$			
			&$0.65 - 18 \,i$
			&$-0.53 - 17 \,i$ 
			&$-12 - 30\, i$ 
			&$-11 + 27\, i$ 
			&$34.3 + 2.4\, i$ \\
			\rule[15pt]{0pt}{0pt}
			$(\delta_{eLR})_{31}\times10^{7}$			
			&$-0.22+0.14\, i$ 
			&$-0.10+0.49\, i$ 
			&$0.62-0.97 \,i$ 
			&$-0.10+1.8 \,i$ 
			&$0.087+0.54 \,i$ \\ 
			\rule[15pt]{0pt}{0pt}
			$(\delta_{eRR})_{21}\times10^{5}$			
			&$-0.04-0.23 \,i$ 
			&$-48+21\, i$ 
			&$0.19-0.009 \,i$  
			&$26+1.4\, i$ 
			&$0$ \\
			\rule[15pt]{0pt}{0pt}
			$(\delta_{eRR})_{31}\times10^{5}$			
			&$-2.99+2.41\, i$ 
			&$-0.82+9.2 \,i$ 
			&$2.2-9.2 \,i$ 
			&$4.5+14 \,i$ 
			&$0$ \\
			\rule[15pt]{0pt}{0pt}
			$(\delta_{eRR})_{32}\times10^{2}$			
			&$0.72+1.3 \,i$ 
			&$0.7-1.2\, i$ 
			&$-0.38+1.2\, i$ 
			&$0.39+1.3\, i$ 
			&$0$\\
			\rule[15pt]{0pt}{0pt}
			flavor-conserving $d_e/e$\,cm 
			&$7.2 \times 10^{-31}$ 
			&$-3.7 \times 10^{-30}$  
			&$1.3 \times 10^{-30}$ 
			&$7.6 \times 10^{-30}$
			&$2.9 \times 10^{-30}$ \\ 
			\rule[15pt]{0pt}{0pt}
			flavored $d_e/e$\,cm 
			&$-5.1 \times 10^{-32}$ 
			&$-2.4\times 10^{-31}$ 
			&$-3.9 \times 10^{-31}$ 
			&$2.5 \times 10^{-31}$
			&$-1.9 \times 10^{-37}$\\ 
			\rule[15pt]{0pt}{0pt}
			$|\sum d_e/e|$\,cm 
			&$6.7\times 10^{-31}$
			&$3.9 \times 10^{-30}$   
			&$9.1 \times 10^{-31}$ 
			&$7.9 \times 10^{-30}$ 
			&$2.9 \times 10^{-30}$\\ 
			\rule[15pt]{0pt}{0pt}
			${\rm BR}(\mu\to e\gamma)$
			&$7.8 \times 10^{-15}$
			&$6.6 \times 10^{-15}$ 
			&$4.5 \times 10^{-14}$ 
			&$4.8 \times 10^{-14}$ 
			&$1.4 \times 10^{-13}$  \\
			\rule[15pt]{0pt}{0pt}
			$	{\rm BR}(\mu\rightarrow   e e \bar e)$&$4.6\times 10^{-17}$  & $3.9\times 10^{-17}$& $2.7\times 10^{-16}$&$2.8\times 10^{-16}$&$8.3\times 10^{-16}$\\ 
			\rule[15pt]{0pt}{0pt}
			${\rm CR}(\mu N\to e N)$&$5.7\times 10^{-17}$ &$4.8\times 10^{-17}$ &$3.3\times 10^{-16}$ &$3.5\times 10^{-16}$ &$1.0\times 10^{-15}$ \\ 
			\hline
		\end{tabular}
	}
	\caption{
		Mass insertion parameters
		and predicted  $d_e/e$
		(flavor-conserving EDM, flavored one and those sum), 
		${\rm BR}(\mu\to e \gamma)$,
		${\rm BR}(\mu\rightarrow  e e \bar e)$ and 
		${\rm CR}(\mu N\to e N)$ 
		at $M_{1/2}=6$\,TeV,  {$m_0=10$\,TeV and $F^\tau=5$\,TeV.} 
	}
	\label{tb:summary}
\end{table}


We have listed  predicted mass insertion parameters
 and  $d_e/e$, ${\rm BR}(\mu\to e \gamma)$,
 ${\rm BR}(\mu\rightarrow  e e \bar e)$ and
 ${\rm CR}(\mu N\to e N)$  
 at $M_{1/2}=6$\,TeV, $m_0=10$\,TeV and  $F^\tau=5$\,TeV
 in Table \ref{tb:summary}.
In this setup, the gaugino masses are given as $M_1=2.9$\,TeV and $M_2=5.2$\,TeV
 at $Q=1$\,TeV, respectively.


The dominant contribution to the electron EDM comes from 
 the imaginary part of the single chirality flipping diagonal mass insertion $(\delta_{eLR})_{11}$ (flavor-conserving EDM) as seen in Eq.\,(\ref{EDM}). 
Therefore, the flavor-conserving EDM
is  almost proportional to  ${\rm Im }(\delta_{eLR})_{11}$
up to its sign. The small differences of  $\tilde m_{eR}$
 among five cases cause   the slight dispersion of the proportionality.
 The largest flavor-conserving EDM is obtained in case D.
The next-to-leading term is the flavored EDM,
which  arises mainly  from  the non-vanishing $(\delta_{eRR})_{ij}$.
 Therefore, it is considerably suppressed in case E
  since  $(\delta_{eRR})_{ij}$ vanish.
  The  largest magnitude of the  flavored EDM is obtained in case C.
  The sum of flavor-conserving EDM and flavored one
  is in the range of  $(0.62$--$7.2)\times 10^{-30}$\,cm for all cases.
  The future  experiment can  reach up to this range.

We can also calculate the leading contribution
of the muon EDM by using  $(\delta_{eLR})_{22}$ of Table.\,\ref{tb:summary}. It is predicted as: 
\begin{equation}
|d_\mu/ e| \simeq 5 \times 10^{-27} \, {\rm cm}\, ,
\end{equation}
for case A.  This predicted value  is significantly below its observed upper bound, 
$2\times 10^{-19}$ at BNL-E821 \cite{Griffith}.
The improvement up  to   $2\times 10^{-21}$ is expected  at FNAL \cite{Griffith}.

The leading terms of
the branching ratio  ${\rm BR}(\mu\to e \gamma)$ are
given in terms of 
$(\delta_{eLR})_{12}$ and $(\delta_{eLR})_{12}^*$ as seen in 
Eq.\,(\ref{BRmu}) due to the chiral enhancement.
The next-to-leading ones arise from $(\delta_{eRR})_{ij}$.
However,  the contribution of the next-to-leading terms
 are  suppressed enough compared with
 the leading ones in all cases.
 The  branching ratio is predicted 
 in  rather broad range  
 $6.6\times 10^{-15}$--$9.4\times 10^{-14}$ for all cases.
Case E will be  tested since the future sensitivity is
 expected to be $6\times 10^{-14}$ \cite{Baldini:2018nnn}.


In SUSY models, the branching ratio of $\mu\to 3e$ and the conversion rate of $\mu N \to e N$  are simply 	related to 
 ${\rm BR}(\mu\to e \gamma)$ as seen in Eq.\,(\ref{CR}).
The five branching ratio and conversion rate 
 are enough below the current experimental upper bounds
 $1.0\times 10^{-12}$ and  $7.0\times 10^{-13}$  \cite{Zyla:2020zbs},
 respectively, 
 as seen in Table \ref{tb:summary}.
 Since future experiments will explore these predictions 
 at the level of $10^{-16}$ for
 $\mu\to 3e$ and  $\mu N \to e N$
 \cite{Blondel:2013ia,Abrams:2012er, Adamov:2018vin},
 it will probe the SUSY mass scale  of $m_0\simeq 10$\,TeV.


We can also calculate the branching ratios of tauon decays, 
$\tau\rightarrow e \gamma$ and $\tau\rightarrow \mu  \gamma$.
Both  branching ratios are
at most ${\cal O}(10^{-13})$, which are  much  below the current experimental upper bounds  $3.3\times 10^{-8}$ and $4.4\times 10^{-8}$,
respectively \cite{Zyla:2020zbs}. 

As well known,  large flavor-violating trilinear coupling may generate instabilities of the electroweak vacuum,  which constrains
the magnitudes of  mass insertion parameters $\delta_{eLR}$ \cite{Park:2010wf}.
It is noted that our predicted ones  do not spoil the vacuum stability.

We also comment on  the effect of neutrino Yukawa matrix in 
the type-I seesaw to the  $\mu\to e \gamma$ decay.
If there are right-handed neutrinos which  couple to the
left-handed neutrinos via Yukawa couplings, the RGEs effects,
 which is  the  running
 from the high scale  $Q_0$ to the right-handed Majorana mass scale $M_R$, can also induce off-diagonal elements in the slepton mass matrix
as follows \cite{Hisano:1995nq,Hisano:1995cp}:
\begin{eqnarray}
(\delta_{e LL})_{12}\simeq -\frac{6m_0^2}{16\pi^2 m_{\rm slepton}^2} 
(Y_D^\dagger Y_D)_{12} \ln \frac{Q_0}{M_R}\, ,
\end{eqnarray}
where $Y_D$ is Dirac neutrino Yukawa matrix in the diagonal base of the charged lepton.
One should check its effect to the  $\mu\to e \gamma$ decay
since our models use type-I seesaw. 
In conclusion, the effect of neutrino Yukawa couplings
 is still at the next-to-leading order of our prediction
as far as we take  $M_R\leq 10^{13}$\,GeV.
For example, in case A, we have 
\begin{eqnarray}
(\delta_{eLL})_{12}\simeq 4.9\times 10^{-3}\
(\,6.6\times 10^{-4}\,)\,,  \qquad
{\rm BR}(\mu\to e\gamma)\simeq 4.8\times 10^{-16}\ (\,8.5\times 10^{-18}\,)
\, ,
\end{eqnarray}
for $M_R=10^{13}\,(10^{12})$\,GeV,
where we take $M_1=2.9$\,TeV, $M_2=5.2$\,TeV and $m_{\rm slepton}=m_0=10$\,TeV, and
the branching ratio includes only the contribution of neutrino Yukawa
couplings.

	In our numerical results, we take  $F^\tau=m_0/2$
	 to prevent the tachyonic slepton 
	since the largest weight is $k_\tau=7$ in our lepton mass matrix.
	Our predicted electron EDM is almost proportional to 
	the magnitude of  $F^\tau$, and
	 the branching ratio of $\mu\to e \gamma$
	  is roughly proportional to   $|F^\tau|^2$
	 in the following numerical analyses.
	 We have checked the numerical results in the case of 
	  $F^\tau=m_0$  for  case E,
	  where  tachyonic sleptons are prevented due to small weight  $1$.
	  Indeed, the calculated $|d_e/e|$ is approximately two times lager than  the one for  $F^\tau=m_0/2$,
	  while  ${\rm BR}(\mu\to e \gamma)$ is four  times lager.
	  Thus, we can estimate roughly  $|F^\tau|$ dependence of our numerical results.
\section{Summary}
\label{sec:Summary}

We have studied the electron EDM 
in the supersymmetric $\rm A_4$ modular invariant theory of flavors
with CP invariance. 
The CP symmetry of the lepton sector is broken by fixing modulus $\tau$.
In this framework,
a fixed  $\tau$ also causes the  CP violation in the soft  SUSY breaking terms.  The electron EDM arises from this CP non-conserved soft SUSY breaking terms.
We have examined the electron EDM in the  five cases A--E of charged lepton mass matrices,
which are completely consistent with observed  lepton masses and
PMNS mixing angles.
It is found that the present upper bound of $|d_e/e|\leq 1.1 \times  10^{-29}$
excludes  the SUSY mass scale, $m_0$ and $M_{1/2}$
below $4$--$6$\,TeV depending on  cases A--E.

The SUSY mass scale is also significantly constrained  by {considering} the
experimental upper bound of  the branching ratio of  the $\mu\to e  \gamma$ decay.

In order to see the effect of CP phase in the modulus $\tau$,
we  examine the correlation between 
the electron EDM and the decay rate of the $\mu \rightarrow e \gamma$ decay.
The correlations are clearly seen
in contrast to  models of the conventional flavor symmetry.
We have found the linear correspondence ${\rm BR}(\mu\to e \gamma)$
between  $|d_e/e|$ in the  logarithmic coordinates
for cases   A--E.
The branching ratio is approximately proportional to the square of $|d_e/e|$.
The slope of the line is independent of the value of $F^\tau$ although $F^\tau=m_0/2$ is taken in our calculations.

The predicted $d_e/e$ and ${\rm BR}(\mu\to e \gamma)$
of  case A and case  C are similar each other,  
and  those of case B and case  D are also similar each other.
The $\mu\to e\gamma$ decay  constrains most severely
the SUSY mass scale in case E compared with other cases. 
On the other hand,  the electron EDM constrains most
severely the SUSY mass scale in case D among five cases.

 Although the current experimental upper bound of the electron EDM 
is $|d_e/e|\leq 1.1\times 10^{-29}$\,cm, 
the future sensitivity of the electron EDM is
expected to reach up to $|d_e/e|\simeq 10^{-30}$\,cm. 
Then, the SUSY mass scale will be  significantly constrained 
by $|d_e/e|$.
 Indeed,  it will probe
 the SUSY mass scale of $10$--$20$\,TeV.

On the other hand, 
the future sensitivity of  the branching ratio ${\rm BR}(\mu\to e \gamma)$,  $6\times 10^{-14}$ 
 probes at most the SUSY mass scale  of  $10$\,TeV.
It is also remarked that
the branching ratio of $\mu\to 3e$ and the conversion rate of $\mu N \to e N$  will be  sensitive for probing the SUSY mass scale of higher than $10$\,TeV.

Thus,
the electron EDM  provides a severe test of the CP violation via the modulus $\tau$ in the supersymmetric modular invariant theory of flavors.

\subsubsection*{Acknowledgments}

We thank  Tatsuo Kobayashi for useful discussions.
The work of K.Y. was supported by the JSPS KAKENHI 21K13923.

\appendix
\section*{Appendix}

\section{Tensor product of  $\rm A_4$ group}
\label{app:A4}
We take the generators of $\rm A_4$ group for the triplet as follows:
\begin{align}
\begin{aligned}
S=\frac{1}{3}
\begin{pmatrix}
-1 & 2 & 2 \\
2 &-1 & 2 \\
2 & 2 &-1
\end{pmatrix},
\end{aligned}
\qquad 
\begin{aligned}
T=
\begin{pmatrix}
1 & 0& 0 \\
0 &\omega& 0 \\
0 & 0 & \omega^2
\end{pmatrix}, 
\end{aligned}
\label{STbase}
\end{align}
where $\omega=e^{i\frac{2}{3}\pi}$ for a triplet.
In this base,
the multiplication rule is
\begin{align}
\begin{pmatrix}
a_1\\
a_2\\
a_3
\end{pmatrix}_{\bf 3}
\otimes 
\begin{pmatrix}
b_1\\
b_2\\
b_3
\end{pmatrix}_{\bf 3}
&=\left (a_1b_1+a_2b_3+a_3b_2\right )_{\bf 1} 
\oplus \left (a_3b_3+a_1b_2+a_2b_1\right )_{{\bf 1}'} \nonumber \\
& \oplus \left (a_2b_2+a_1b_3+a_3b_1\right )_{{\bf 1}''} \nonumber \\
&\oplus \frac13
\begin{pmatrix}
2a_1b_1-a_2b_3-a_3b_2 \\
2a_3b_3-a_1b_2-a_2b_1 \\
2a_2b_2-a_1b_3-a_3b_1
\end{pmatrix}_{{\bf 3}}
\oplus \frac12
\begin{pmatrix}
a_2b_3-a_3b_2 \\
a_1b_2-a_2b_1 \\
a_3b_1-a_1b_3
\end{pmatrix}_{{\bf 3}\  } \ , \nonumber \\
\nonumber \\
{\bf 1} \otimes {\bf 1} = {\bf 1} \ , \qquad &
{\bf 1'} \otimes {\bf 1'} = {\bf 1''} \ , \qquad
{\bf 1''} \otimes {\bf 1''} = {\bf 1'} \ , \qquad
{\bf 1'} \otimes {\bf 1''} = {\bf 1} \  ,
\end{align}
where
\begin{align}
T({\bf 1')}=\omega\,,\qquad T({\bf 1''})=\omega^2. 
\end{align}
More details are shown in the review~\cite{Ishimori:2010au,Ishimori:2012zz}.

\section{Modular forms with weight 2,\, 4,\, 6,\, 8 in $\Gamma_3$ group}
\label{app:modular}
We present modular forms  with weight 2,\, 4,\, 6,\, 8
in $\rm A_4$ modular group.
 The triplet modular forms can be written in terms of 
$\eta(\tau)$ and its derivative \cite{Feruglio:2017spp}:
\begin{eqnarray} 
\label{eq:Y-A40}
Y_1 &=& \frac{i}{2\pi}\left( \frac{\eta'(\tau/3)}{\eta(\tau/3)}  +\frac{\eta'((\tau +1)/3)}{\eta((\tau+1)/3)}  
+\frac{\eta'((\tau +2)/3)}{\eta((\tau+2)/3)} - \frac{27\eta'(3\tau)}{\eta(3\tau)}  \right), \nonumber \\
Y_2 &=& \frac{-i}{\pi}\left( \frac{\eta'(\tau/3)}{\eta(\tau/3)}  +\omega^2\frac{\eta'((\tau +1)/3)}{\eta((\tau+1)/3)}  
+\omega \frac{\eta'((\tau +2)/3)}{\eta((\tau+2)/3)}  \right) , \label{Yi} \\ 
Y_3 &=& \frac{-i}{\pi}\left( \frac{\eta'(\tau/3)}{\eta(\tau/3)}  +\omega\frac{\eta'((\tau +1)/3)}{\eta((\tau+1)/3)}  
+\omega^2 \frac{\eta'((\tau +2)/3)}{\eta((\tau+2)/3)}  \right)\,.
\nonumber
\end{eqnarray}
They are also  expressed in the $q$ expansions,
 where $q=e^{i2\pi \tau}$, as follows:
\begin{align}
{\bf Y^{\rm (2)}_3}(\tau)
=\begin{pmatrix}Y_1\\Y_2\\Y_3\end{pmatrix}=
\begin{pmatrix}
1+12q+36q^2+12q^3+\dots \\
-6q^{1/3}(1+7q+8q^2+\dots) \\
-18q^{2/3}(1+2q+5q^2+\dots)\end{pmatrix}.
\label{Y(2)}
\end{align}
For weight 4,  there are  five modular forms
by the tensor product of  $\bf 3\otimes 3$ as:
\begin{align}
&\begin{aligned}
{\bf Y^{\rm (4)}_1}(\tau)=Y_1^2+2 Y_2 Y_3 \, , \qquad\quad\ \
{\bf Y^{\rm (4)}_{1'}}(\tau)=Y_3^2+2 Y_1 Y_2 \, , 
\end{aligned}\nonumber \\
\nonumber \\
&\begin{aligned} 
{\bf Y^{\rm (4)}_{3}}(\tau)=
\begin{pmatrix}
Y_1^{(4)} \\
Y_2^{(4)}\\
Y_3^{(4)}
\end{pmatrix}
=
\begin{pmatrix}
Y_1^2-Y_2 Y_3 \\
Y_3^2 -Y_1 Y_2\\
Y_2^2-Y_1 Y_3
\end{pmatrix}\, . 
\end{aligned}
\label{weight4}
\end{align}
For $k=6$, there are  seven modular forms
by the tensor products of  $\rm A_4$ as:
\begin{align}
&\begin{aligned}
{\bf Y^{(\rm 6)}_1}=Y_1^3+ Y_2^3+Y_3^3 -3Y_1 Y_2 Y_3  \ , 
\end{aligned} \nonumber \\
\nonumber \\
&\begin{aligned} {\bf Y^{(\rm 6)}_3}\equiv 
\begin{pmatrix}
Y_1^{(6)}  \\
Y_2^{(6)} \\
Y_3^{(6)}
\end{pmatrix}
=(Y_1^2+2  Y_2 Y_3)
\begin{pmatrix}
Y_1   \\
Y_2 \\
Y_3
\end{pmatrix}\ , \qquad
\end{aligned}
\begin{aligned} {\bf Y^{(\rm 6)}_{3'}}\equiv
\begin{pmatrix}
Y_1^{'(6)}  \\
Y_2^{'(6)} \\
Y_3^{'(6)}
\end{pmatrix}
=(Y_3^2+2 Y_1 Y_2)
\begin{pmatrix}
Y_3   \\
Y_1 \\
Y_2
\end{pmatrix}\ . 
\end{aligned}
\label{weight6}
\end{align}
For $k=8$, there are  9 modular forms
by the tensor products of  $\rm A_4$ as:
\begin{align}
&\begin{aligned}
{\bf Y^{(\rm 8)}_1}=(Y_1^2+ 2 Y_2 Y_3)^2 \, ,\qquad  
{\bf Y^{(\rm 8)}_{1'}}=(Y_1^2+ 2 Y_2 Y_3)(Y_3^2+ 2 Y_1 Y_2)\, ,  \qquad
{\bf Y^{(\rm 8)}_{1"}}=(Y_3^2+ 2 Y_1 Y_2)^2 \, ,
\end{aligned} \nonumber \\
&\begin{aligned} {\bf Y^{(\rm 8)}_3}\equiv 
\begin{pmatrix}
Y_1^{(8)}  \\
Y_2^{(8)} \\
Y_3^{(8)}
\end{pmatrix}
= (Y_1^3+Y_2^3+Y_3^3-3 Y_1 Y_2 Y_3 )
\begin{pmatrix}
Y_1 \\
Y_2 \\
Y_3
\end{pmatrix} , \ 
\end{aligned}
\begin{aligned} {\bf Y^{(\rm 8)}_{3'}}\equiv
\begin{pmatrix}
Y_1^{'(8)}  \\
Y_2^{'(8)} \\
Y_3^{'(8)}
\end{pmatrix}
=(Y_3^2+ 2 Y_1 Y_2)
\begin{pmatrix}
Y_2^2 - Y_1  Y_3   \\
Y_1^2 - Y_2  Y_3  \\
Y_3^2 - Y_1  Y_2 
\end{pmatrix} . \nonumber
\end{aligned}
\label{weight8}
\end{align}

\section{RGEs of leptons and slepton}
\label{app:RGE}

The relevant RGEs  are given by 
\cite{Martin:1993zk,Martin:1997ns};
\begin{eqnarray}
\begin{split}
16\pi^2 \frac{d}{d t} \left( {\tilde m}^2_{eL} \right)_{ij}
=&   -\left( \frac{6}{5} g_1^2 \left| M_1 \right|^2
+ 6 g_2^2 \left| M_2 \right|^2 \right) \delta_{ij}
-\frac{3}{5} g_1^2~S~\delta_{ij}
\\
&+  \left (( {\tilde m}^2_{eL} ) {Y}_e^{\dagger} {Y}_e
+ {Y}_e^{\dagger} {Y}_e ( {\tilde m}^2_{eL} )_{K} \right)_{ij} 
\\
&+ 2 \left( {Y}_e^{\dagger} ({\tilde m}^2_{eR} )_{K} {Y}_e
+{\tilde m}^2_{H_d} {Y}_e^{\dagger} {Y}_e
+{A}_e^{\dagger} {A}_e \right)_{ij} \ ,
\\
16\pi^2 \frac{d}{d t} \left( {\tilde m}^2_{eR} \right)_{ij} =&
- \frac{24}{5} g_1^2 \left| M_1 \right|^2 \delta_{ij} + \frac{6}{5} g_1^2~S~\delta_{ij} 
\\
&+ 2 \left ( ( {\tilde m}^2_{eR} )_{K}  {Y}_e {Y}_e^{\dagger} 
+ {Y}_e {Y}_e^{\dagger} ( {\tilde m}^2_{eR} )_{K} \right)_{ij}
\\
\label{RGE}
&+ 4 \left( {Y}_e ( {\tilde m}^2_{eL} )_{K} {Y}_e^{\dagger} + {m}^2_{H_d}
{Y}_e {Y}_e^{\dagger} +  {A}_e {A}_e^{\dagger} \right)_{ij}~, 
\\
16\pi^2 \frac{d}{d t}  \left( {A}_e \right)_{ij} =&
\left( -\frac{9}{5} g_1^2 -3 g_2^2
+ 3 {\rm Tr} ( {Y}_d^{\dagger} {Y}_d )
+   {\rm Tr} ( {Y}_e^{\dagger} {Y}_e ) \right )  \left({A}_{e}\right)_{ij} 
\\
&+ 2 \left(
\frac{9}{5} g_1^2 M_1 + 3 g_2^2 M_2
+ 3 {\rm Tr} ( {Y}_d^{\dagger} {A}_d)
+   {\rm Tr} ( {Y}_e^{\dagger} {A}_e) \right) {Y}_{e_{ij}} 
\\
&+ 4 \left( {Y}_e {Y}_e^{\dagger} {A}_e \right)_{ij}
+ 5 \left({A}_e {Y}_e^{\dagger} {Y}_e \right)_{ij}~, 
\\
16\pi^2 \frac{d}{d t} {Y}_{e_{ij}} =& \left ( -\frac{9}{5} g_1^2 - 3 g_2^2
+ 3 \,{\rm Tr} ( {Y}_d {Y}_d^{\dagger})
+   {\rm Tr} ( {Y}_e {Y}_e^{\dagger})
\right ) {Y}_{e_{ij}}
+ 3 \, \left( {Y}_e {Y}_e^{\dagger} {Y}_e \right)_{ij}\ .
\end{split}
\label{RGE}
\end{eqnarray}
In these expressions,  $g_{1,2}$ are the gauge couplings of 
SU(2)$_L\times U(1)_Y$,  
$M_{1,2}$ are the corresponding gaugino mass terms, 
${Y}_{e,d}$ 
are the Yukawa couplings for charged leptons and down quarks, 
${A}_e= ({\tilde m}^2_{eLR})/v_d$,  and
\begin{equation}
S = {\rm Tr} ({\tilde m}^2_{qL} + {\tilde m}^2_{dR}- 2 {\tilde m}^2_{uR}
- {\tilde m}^2_{eL} + {\tilde m}^2_{eR} ) - {\tilde m}^2_{H_d}
+ {\tilde m}^2_{H_u} \nonumber,
\end{equation}
where ${\tilde m}^2_{qL}$, ${\tilde m}^2_{dL}$, ${\tilde m}^2_{uR}$ are mass matrices of 
squarks and $\tilde m_{H_u}$ and $\tilde m_{H_d}$  are the Higgs  masses. 
The parameter $t$ is $t=\ln Q/Q_0$, where $Q$ is the renormalization scale
and $Q_0$ is a reference scale.


\section{Loop functions}
\label{app:Loopfun}

The dimensionless functions $C_{B}$, $C'_{B,R}$,
$C'_{B,L}$ and $C''_B$ are given approximately   as \cite{Dimou:2015cmw}: 
\begin{eqnarray}
C_B  & \simeq  & 
h_1(\bar x) \, ,\\
C'_{B,R}&\simeq&C'_{B,L}\simeq\frac{1}{2}\,[h_1(\bar x)+k_1(\bar x)]\,,\\
C''_B & \simeq  & \frac{1}{3}\, [h_1(\bar x)+2\, k_1(\bar x)]\, ,\\
\end{eqnarray}
where
we take $\tilde m_{e}=\sqrt{\tilde m_{eL}\tilde m_{eR}}$
as the average slepton mass
and put $\bar{x}=(M_1/\tilde m_{e})^2$.

On the other hand,
functions $C'_{L}$, $C'_{R}$ and 
$C'_{2}$  are exactly  given  as: 
\begin{eqnarray}
C'_L & =  &C_0 \frac{1}{m_{eL}^2}
\frac{y_L}{y_L - x_L}\left[\, h_1 \left( x_L\right)
- \, h_1 \left( y_L\right)\right] ,\\ 
C'_R & =  & C_0\frac{1}{m_{eR}^2}
\frac{y_R}{y_R - x_R}\left[\, h_1 \left( x_R\right) 
- \, h_1 \left( y_R\right)\right] , \\
C'_2 & = & C_0\frac{M_2\cot ^2\theta_W}{M_1 m_{eL}^2}
\frac{y_L}{y_L - x'_L} \left[\, h_2 \left( x'_L\right) 
- \, h_2 \left( y_L\right)\right] , \\ 
\label{Iapp}
\end{eqnarray}
with
\begin{eqnarray}
x_L= \frac{M_1^2}{m_{eL}^2}\,,~~\quad x_R= \frac{M_1^2}{m_{eR}^2}\, ,~~\quad x'_L=
\frac{M_2^2}{m_{eL}^2}\, ,~~ \quad y_L= \frac{\mu^2}{m_{eL}^2}\,,~~\quad y_R= \frac{\mu^2}{m_{eR}^2}\, ,
\end{eqnarray}
where $C_0=m_0^4/\mu^2$.
Functions $h_1 (x)$,  $h_2 (x)$ and  $k_1 (x)$ are defined as:
\begin{eqnarray}
h_1 (x) &=& \frac{1+4x-5x^2 + (2x^2 + 4x)\ln x}{(1 - x)^4}\, ,\\
h_2 (x) &=& \frac{7x^2+4x-11 - 2(x^2 +6x+2) \ln x }{2(x - 1)^4} \, ,\\
k_1(x)&=&\frac{d}{dx} [x\, h_1(x)]\, .
\end{eqnarray}
Note that  $M_i$ and $\mu^2$ are real positive values.

\section{ Mass matrix $M_e$ with only weight 2 modular forms}
\label{app:Me}

We present the charged lepton mass matrix  in terms of only
weight 2 modular forms,
 in which  ${\rm BR}(\mu\to e \gamma)$
  was discussed in Ref. \cite{Kobayashi:2021bgy}.
The mass matrix is given as:
\begin{align}
\begin{aligned}
M_e=v_d
\begin{pmatrix}
\alpha_e& 0 & 0 \\
0 & \beta_e &0 \\
0 &0& \gamma_e
\end{pmatrix}
\begin{pmatrix}
Y_1 & Y_3 & Y_2 \\
Y_2 & Y_1 & Y_3 \\
Y_3 & Y_2 & Y_1
\end{pmatrix}\,,
\end{aligned}
\end{align}
where $Y_i$'s are given in  Eq.\,(\ref{eq:Y-A40}).
The neutrino mass matrix is given by the dimension five Weinberg operator.
We present the best fit parameter set for
the observed lepton masses and flavor mixing angles \cite{Okada:2020brs} as follows:
\begin{equation}
E\,: \ \  \tau=-0.0796 + 1.0065  \, i \,, \qquad  
\alpha_e/\gamma_e=6.82\times 10^{-2}\,, \qquad 
\beta_e/\gamma_e=1.02\times 10^{-3}\,,
\label{tau0}
\end{equation}
which is referred as the case E in the text.
 
The slepton mass matrix $\tilde m^2_{eRL}$ 
including RGE effect is written as:
\begin{align}
\tilde m^2_{eRL}\simeq 1.4\,v_d \left[- m_F
\begin{pmatrix}
\alpha_e & 0 & 0 \\
0 &\beta_e & 0\\
0 & 0 &\gamma_e
\end{pmatrix} 
\begin{pmatrix}
Y_1 & Y_3 &  Y_2 \\
Y_2 &  Y_1&  Y_3\\
Y_3 & Y_2& Y_1 \\
\end{pmatrix}
+  F^\tau
\begin{pmatrix}
\alpha_e & 0 & 0 \\
0 &\beta_e & 0\\
0 & 0 &\gamma_e
\end{pmatrix} \frac{d}{d \tau}
\begin{pmatrix}
Y_1 & Y_3 &  Y_2\\
Y_2 &  Y_1&  Y_3\\
Y_3 & Y_2& Y_1 \\
\end{pmatrix}\right ]\,.
\end{align}
Since the first term of the right-hand side
 is proportional to $M_e$, it does not contributes
  to  $|d_e/e|$ and ${\rm BR}(\mu\to e \gamma)$.
  On the other hand, $\tilde m^2_{eLL}$ and $\tilde m^2_{eRR}$ 
  are proportional to the unit matrix. Then,
   these mass terms do not contribute to
   $|d_e/e|$ and ${\rm BR}(\mu\to e \gamma)$.



\end{document}